\documentclass[11pt]{article}

\usepackage[T1]{fontenc}
\usepackage{lmodern}
\usepackage[margin=1in]{geometry}
\usepackage{amsmath,amssymb,amsthm,mathtools}
\usepackage{microtype}
\usepackage{enumitem}
\usepackage{needspace}
\usepackage{booktabs,tabularx,array}
\usepackage{tikz}
\usetikzlibrary{arrows.meta,positioning,fit,backgrounds,calc}
\usepackage{xurl} 
\usepackage[colorlinks=true,linkcolor=blue,citecolor=blue,urlcolor=blue]{hyperref}
\hypersetup{pdftitle={Optimal-Size Strong Connectivity Preservers under Two Edge Failures in Subcubic Time},pdfauthor={Wanhee Lee and Chenglin Fan}}
\setlist{itemsep=3pt,topsep=5pt}
\newtheorem{theorem}{Theorem}[section]
\newtheorem{lemma}[theorem]{Lemma}
\newtheorem{corollary}[theorem]{Corollary}
\newtheorem{proposition}[theorem]{Proposition}
\theoremstyle{definition}

\newtheorem{algorithm}{Algorithm}
\theoremstyle{remark}

\newcommand{\SCC}{\operatorname{SCC}}
\newcommand{\flow}{\operatorname{flow}}
\newcommand{\polylog}{\operatorname{polylog}}
\newcommand{\softO}{\widetilde O}
\newcommand{\F}{\mathbb F}
\newcommand{\rev}{\mathrm{rev}}
\newcommand{\Local}{\textnormal{\textsc{LocalSix}}}

\newcommand{\transpose}{\mathsf T}
\newcommand{\FAIL}{\textnormal{\textsc{Fail}}}
\newcommand{\basefield}{\mathbb F_{q_0}}

\title{Optimal-Size Strong Connectivity Preservers\\
under Two Edge Failures in Subcubic Time}
\author{Wanhee Lee\thanks{Department of Computer Science and Engineering,
Seoul National University, Seoul 08826, South Korea.
\texttt{wanhee04@snu.ac.kr}}
\and Chenglin Fan\thanks{Department of Computer Science and Engineering,
Seoul National University, Seoul 08826, South Korea.
\texttt{fanchenglin@snu.ac.kr}}}
\date{}
\begin{document}
\begin{titlepage}
\maketitle
\thispagestyle{empty}
\begin{abstract}
A $k$-edge-failure strong connectivity preserver of a directed graph on
$N$ vertices is a spanning subgraph that has the same strongly connected
components as the input after each set of at most $k$ edge deletions.
The lower-bound construction of Baswana, Choudhary, and
Roditty~\cite{bcr}, adapted to strong connectivity in~\cite{cc}, gives the previous
best bound of $\Omega(2^kN)$ necessary edges. Hoppenworth, Saranurak,
and Wang~\cite{hsw} prove an $O(k4^kN\log N)$ upper bound, leaving a
logarithmic gap for fixed $k$. We close this gap by giving an explicit
strongly connected graph on $N=2^d$ vertices with
$N-1+(N/2)\log_2N$ edges, every one of which is necessary in a
two-failure preserver. Together with the upper bound of~\cite{hsw},
this proves that the optimal size in the worst case is $\Theta_k(N\log N)$
for every fixed $k\geq2$, whereas it is $\Theta(N)$ for $k=1$~\cite{gip}.
The same graph has a one-failure preserver with at most $3(N-1)$ edges.
We also give a randomized algorithm that either returns a spanning
subgraph or reports failure. For every fixed $c,\varepsilon>0$, with probability
at least $1-N^{-c}$ it returns a correct preserver for $k=2$ with
$O(N\log N)$ edges, in $\softO(N^{(\omega+3)/2+\varepsilon})$ time and
$\softO(N^2)$ space. Here $\omega$ is the matrix multiplication
exponent over a fixed finite field.
\end{abstract}
\end{titlepage}
\setcounter{page}{1}

\section{Introduction}\label{sec:intro}

Deleting a few edges can split a strongly connected component into many
smaller components. A sparse subgraph that preserves these splits gives
one compact representation for all possible failure sets. The required
subgraph must retain original edges, and it must work for every failure
set at once. Preserving only whether the whole graph remains strongly
connected is a weaker task.

Let $G=(V,E)$ be a simple directed graph with $N=|V|$ vertices: it has
no parallel edges or self-loops, but opposite edges are allowed. A
strongly connected component (SCC) is a maximal vertex set whose
vertices can all reach one another; $\SCC(X)$ denotes the partition of
$V(X)$ into these sets. A \emph{$k$-edge-failure SCC preserver}, also
called an SCC certificate, is a spanning subgraph $H\subseteq G$
(one keeping every vertex) such that
\begin{equation}\label{eq:definition}
 \SCC(H-F)=\SCC(G-F)
 \qquad(F\subseteq E,\ |F|\leq k).
\end{equation}
Deleting an edge that is absent from $H$ has no effect on $H$.
We study both the number of retained edges and the time needed to select
them. All failures in this paper are edge failures. We call an SCC of $G-F$
a \emph{failure SCC}.

\subsection{Prior bounds and the questions they leave}

For one failure, Georgiadis, Italiano, and Parotsidis~\cite{gip} give an
$O(N)$-edge certificate in $O(|E|+N)$ time. For a fixed root $s$, a \emph{rooted preserver} retains reachability
from $s$ to every vertex after each allowed failure set. Baswana,
Choudhary, and Roditty (BCR)~\cite[Theorem 2]{bcr} construct such a
preserver for $k$ failures with at most $2^kN$ edges in
$O(2^k|E|N)$ time. An SCC preserver must also protect components that
do not contain the chosen root, so this rooted guarantee alone does not
give a linear-size solution to~\eqref{eq:definition}.

Chakraborty and Choudhary~\cite{cc} proved the size bound
$\softO(k2^kN^{2-1/k})$ for SCC preservers. Hoppenworth, Saranurak,
and Wang (HSW)~\cite[Theorems 1.1 and 4.2]{hsw} then proved the existence of
$O(k4^kN\log N)$-edge preservers and described an $N^{O(k)}$ construction.
Their separate construction with $O(8^kN\log^{5/2}N)$ edges runs in
$\operatorname{poly}(2^kN)$ time~\cite[Theorem 4.12]{hsw}.
The previous best lower bound, $\Omega(2^kN)$, comes from the construction
of Baswana, Choudhary, and Roditty~\cite[Theorem 3 and Section 7]{bcr}.
Their theorem concerns single-source reachability. Chakraborty and
Choudhary~\cite[Appendix A]{cc} give the SCC adaptation by adding
return edges; HSW also describe this construction~\cite[Theorem B.3]{hsw}.\footnote{Their graph consists of
an outward binary tree of height $k$, rooted at $s$, with leaf set $X$,
and a disjoint set $Y$, with all edges from $X$ to $Y$.
Add an edge $y\to s$ for every $y\in Y$.
For each $x\in X$, failing the $k$ sibling edges along the root-to-$x$
path makes $x$ the only leaf reachable from $s$.
Each edge $x\to y$ is then necessary to keep $s$ and $y$ in the same SCC.
When $N\geq3\cdot2^k$, these edges number at least $2^kN/3$.}
HSW leave the logarithmic gap open~\cite[Section 7]{hsw}.
For $k=2$, these results raise two questions: is $N\log N$ necessary,
and can a preserver of that size be constructed in subcubic time?

We answer both questions. Table~\ref{tab:prior} separates stated published
results from a baseline derived here. Protecting a linear-size
one-failure certificate~\cite{gip} and using the HSW size
theorem~\cite[Theorem 4.2]{hsw} already gives an $O(N^5)$ construction. Appendix~\ref{sec:baseline}
proves this composition. The construction below achieves subcubic time
while retaining the $O(N\log N)$ size bound.

\begin{table}[t]
\centering\small
\begin{tabularx}{\textwidth}{@{}p{0.34\textwidth}p{0.25\textwidth}X@{}}
\toprule
Result and $k$ & Output edges & Construction time\\
\midrule
GIP~\cite{gip}, $k=1$ & $O(N)$ & $O(|E|+N)$\\
CC~\cite{cc}, $k=2$ & $\softO(N^{3/2})$ & Polynomial\\
HSW~\cite{hsw}, $k=2$, first bound & $O(N\log N)$ & $N^{O(1)}$\\
HSW~\cite{hsw}, $k=2$, second bound & $O(N\log^{5/2}N)$ & Polynomial\\
Derived baseline, Proposition~\ref{prop:baseline} & $O(N\log N)$ & $O(N^5)$, deterministic\\
This paper, Theorem~\ref{thm:main} & $O(N\log N)$ & $\softO(N^{(\omega+3)/2+\varepsilon})$, randomized\\
\bottomrule
\end{tabularx}
\caption{All rows preserve the full SCC partition for every allowed edge
failure set. Appendix~\ref{sec:baseline} derives the baseline from the GIP
certificate and the HSW size theorem. In the last row, $\varepsilon>0$
is arbitrary and fixed.}
\label{tab:prior}
\end{table}

\subsection{Our results}

Fix a finite base field $\basefield$, and let $\omega=\omega_{\basefield}$
be its matrix multiplication exponent, in the bilinear model specified
in Section~\ref{sec:model}.
Throughout the algorithm analysis, soft-$O$ suppresses powers of the
global $\log(N+2)$, where $N$ is the number of vertices in the original input, including
in bounds for a recursion node with $n$ vertices.

\begin{theorem}[Construction]\label{thm:main}
Fix $c>0$, $\varepsilon>0$, and a finite base field $\basefield$.
On a simple directed graph with $N\geq2$ vertices, a randomized
algorithm terminates and returns either $\FAIL$ or a spanning subgraph.
With probability at least $1-N^{-c}$, it returns a two-edge-failure SCC
preserver with $O(N\log(N+2))$ edges in
\[
 \softO\bigl(N^{(\omega+3)/2+\varepsilon}\bigr)
\]
time and $\softO(N^2)$ space. The constants may depend on $c$,
$\varepsilon$, the field, and the chosen matrix algorithm.
\end{theorem}

For the analysis, fix an achievable exponent
$\omega<\omega'<\min\{3,\omega+2\varepsilon\}$ and put
$a=(\omega'+3)/2$. Section~\ref{sec:model} specifies the fixed matrix
algorithm that realizes $\omega'$. We prove the stronger bound
$\softO(N^a)$, with $a<3$, throughout the construction.
The symbol $\FAIL$ reports a detected inconsistency or an exceeded
sampling or test limit. The algorithm is Monte Carlo: an incorrect subgraph may be
returned on the failure event, whose probability also covers any
violation of the stated resource bounds.

\begin{theorem}[Necessary edges]\label{thm:lower-main}
For every integer $d\geq1$, a simple strongly connected directed graph
on $N=2^d$ vertices has
\[
 N-1+\frac N2\log_2N
\]
edges and has no proper two-edge-failure SCC preserver. The same graph
has a one-edge-failure SCC preserver with at most $3(N-1)$ edges.
For every sufficiently large integer $N$, an $N$-vertex strongly
connected graph requires $\Omega(N\log N)$ edges.
\end{theorem}

Let $s_k(N)$ be the largest, over simple $N$-vertex directed graphs, of
the minimum number of edges in a $k$-edge-failure SCC preserver.

\begin{corollary}[Optimal size]\label{cor:optimal-main}
We have $s_1(N)=\Theta(N)$ and, for every fixed integer $k\geq2$,
\[
 s_k(N)=\Theta_k(N\log N).
\]
\end{corollary}

The lower bound uses only two failures and thus applies to every larger
value of $k$. The fixed-$k$ upper bound is HSW's theorem~\cite[Theorem 4.2]{hsw}.
Optimality here concerns the worst-case size $s_k(N)$; an individual
graph can admit a smaller preserver. The subcubic construction applies
to $k=2$, and its optimal running time remains open.

OpenAI~\cite{oai-94} proves $\omega_{\mathbb C}\leq9/4$.
Amanov, van den Berg, and Zuiddam~\cite[Theorem 1.1]{abz} establish
this bound over every field. Taking $\basefield=\mathbb F_2$ in
Theorem~\ref{thm:main} gives a running time of
$\softO(N^{21/8+\varepsilon})$ for every fixed $\varepsilon>0$.

\subsection{Prior techniques and our contributions}

The construction has three main ingredients: HSW's terminal hierarchy~\cite{hsw},
the local reachability-preservation method of Baswana, Choudhary, and
Roditty (BCR)~\cite{bcr}, and the algebraic encoding of Choudhary, Kumar,
and Saggi (CKS)~\cite{cks}. The one-failure certificate and dominator
machinery of Georgiadis, Italiano, and Parotsidis (GIP)~\cite{gip}
support the terminal construction.
Table~\ref{tab:provenance} separates these ingredients from the statements
needed to obtain the present size and running-time bounds.

\begin{table}[t]
\centering\small
\renewcommand{\arraystretch}{1.12}
\begin{tabularx}{\textwidth}{@{}>{\raggedright\arraybackslash}p{0.23\textwidth}>{\raggedright\arraybackslash}X>{\raggedright\arraybackslash}X@{}}
\toprule
Ingredient & Result used from prior work & Adaptation or additional argument\\
\midrule
Local reachability preservation & BCR's farthest-cut method, incoming-edge
selection argument, and locality lemma~\cite[Lemmas 7, 10, and 13]{bcr}. &
The local-call variant with at most two auxiliary shortcuts and its algebraic implementation
(Sections~\ref{sec:local} and~\ref{sec:algebra}).\\[3pt]
Determinant encoding & CKS's common-basis encoding and bounded-flow
reduction~\cite[Sections 3.1 and 4.1]{cks}. &
A circulation variant of the flow-threshold gadget, original boundary-edge
recovery, and the restricted row and column changes used by the queries
(Section~\ref{sec:algebra}).\\[3pt]
Inverse maintenance & Low-rank inverse updates and periodic rebuilding;
see~\cite[Section 2]{bns}. &
Sparse replacement blocks with total retained indegree $O(n)$ give the
pass bound (Section~\ref{sec:batch}).\\[3pt]
One-failure SCC information & GIP's sparse certificate and the
dominator description of a root component~\cite[Theorems 9.1 and 3.7]{gip}. &
A family of $O(n^2)$ representative failure sets, component records, and
updates of terminal counts (Section~\ref{sec:terminals}).\\[3pt]
Balanced terminal hierarchy & HSW's hierarchy and balance-preserving
terminal exchanges~\cite[Appendix C.2]{hsw}. &
At most two terminals outside one failure SCC, with a
$\softO(n^a)$ construction (Section~\ref{sec:terminals}).\\[3pt]
Selected roots and candidate sources & HSW's selected-root and per-vertex
candidate-list construction, using cut containers~\cite[Sections 4.2 and 6]{hsw}. &
Five fixed anchors give $|L_v|\leq40$; two list passes retain
$O(n)$ edges (Sections~\ref{sec:node-passes} and~\ref{sec:lists}).\\
\bottomrule
\end{tabularx}
\caption{Prior results and their use in the construction. Here $n$ is the number
of vertices in the current node graph and $a=(\omega'+3)/2$. The right column records the adaptations and
composition arguments proved in this paper.}
\label{tab:provenance}
\end{table}

A \emph{terminal set} is a designated vertex subset $U$. HSW construct a
balanced hierarchy of such sets and use the concentration of terminals
in one failure SCC to reduce the number of required sources. Their
candidate-list strategy assigns different candidate source sets to different
targets; this principle is part of our starting point.
Their hierarchy construction adapts the undirected hierarchy of
Long, Pettie, and Saranurak~\cite{lps}.
HSW's directed $(q,k)$-unbreakability condition requires that every cut
with at most $k$ edges in either direction has at most $q$ terminals on
one side~\cite[Definition 2.1]{hsw}. It implies that some SCC contains
all but at most $2q$ terminals after $k$ failures~\cite[Claim 3.6]{hsw}.
For $q=k=2$ this permits four exceptions. We construct a balanced set
with at most two exceptions directly. This stronger property supports
the constant-size list construction in Section~\ref{sec:lists}.

The determinant in Section~\ref{sec:algebra} specializes CKS's
algebraic framework: one partition matroid groups edges by
tail, the other by head, and their $K$-fold unions impose the two degree
bounds. We use the resulting common-basis polynomial and a circulation
variant of CKS's bounded-flow reduction~\cite[Section 4.1]{cks}.
The local calls also need to recover original edge identifiers and to maintain the
encoding through a sequence of incoming restrictions that may delete many edges.
Sections~\ref{sec:algebra} and~\ref{sec:batch} supply this query and
update interface, using standard low-rank inverse maintenance and
periodic rebuilding~\cite{san,bns}.

For the terminal construction, GIP's certificate reduces all relevant
two-failure partitions to $O(n^2)$ representatives. Probe graphs, each the union of two rooted preservers,
then allow GIP's one-failure dominator description to be used after
fixing the first failure. The additional task is to maintain terminal
counts in these fixed component records as $U$ changes. The terminal exchange
follows HSW's hierarchy construction. We prove that it yields terminal
concentration and bounds the number of membership changes; the terminal
monitor then gives the stated running time.

Related oracles preprocess the graph to answer queries after failures.
These include BCR's
SCC reporting structure~\cite{bcr-oracle}, strong connectivity oracles
for two vertex failures~\cite{gkt}, algebraic reachability sensitivity
oracles~\cite{vdbs}, and flow sensitivity oracles~\cite{ahi,cks}.
For every constant $k$, Anand, Saranurak, and Wang~\cite{asw} give
a randomized oracle with $N^{1+o(1)}$ space and $N^{o(1)}$ query time
that decides, with high probability, whether the whole graph remains
strongly connected after at most $k$ edge failures.

Bounded-flow preservers~\cite{bounded-flow} retain subgraphs preserving
flow values from a fixed source up to a specified threshold.
For a specified pair set $P$, Chakraborty, Chatterjee, and
Choudhary~\cite{ccc} construct a two-failure reachability preserver
with $O(N^{4/3}|P|^{1/3})$ edges and prove a worst-case lower bound
of $\Omega(N\sqrt{|P|})$ edges, strengthening the earlier pairwise
lower bounds of~\cite{cc}. This pairwise guarantee preserves one-way
reachability even between different failure SCCs. Georgiadis
et al.~\cite{gkkpp} compute $(k+2)$-edge-connected components in
$k$-edge-connected digraphs. These components group vertex pairs that
remain strongly connected after every set of fewer than $k+2$ edge
failures; they do not specify the SCC partition after each individual
failure set.

\section{Technical overview}\label{sec:overview}

\begin{figure}[t]
\centering
\begin{tikzpicture}[>=Latex,font=\small,
  box/.style={draw,rounded corners,align=center,text width=4.0cm,
              minimum height=1.05cm,fill=blue!3},
  impl/.style={->,very thick},logic/.style={->,dashed},
  lab/.style={font=\footnotesize,align=center,inner sep=2pt}]
 \node[box] (u) at (0,0) {Balanced terminal set $U$\\Section~\ref{sec:terminals}};
 \node[box] (j) at (-3.65,-2.25) {Three-root graph $J$\\Section~\ref{sec:node-passes}};
 \node[box] (l) at (3.65,-2.25) {Source lists $L_v$\\$|L_v|\leq40$, Section~\ref{sec:lists}};
 \node[box] (p) at (0,-4.8) {Forward and reverse\\list passes\\Section~\ref{sec:node-passes}};
 \node[box] (d) at (4.1,-7.0) {Recurse on $G[D]$\\$D\in\SCC(G-U)$};
 \node[box] (h) at (0,-9.1) {Union of node\\and child edges};
 \draw[impl] (u) -- (j);
 \draw[impl] (u) -- (l);
 \draw[impl] (j) -- (p);
 \node[lab] at (-3.75,-3.55) {protected\\edges};
 \draw[impl] (l) -- (p);
 \node[lab] at (3.75,-3.55) {sources for\\local calls};
 \draw[logic] (u) -- (p);
 \node[lab,fill=white] at (0,-2.25) {terminal\\concentration};
 \draw[impl] (u.east) -- (7.25,0) -- (7.25,-7.0) -- (d.east);
 \node[lab,rotate=90] at (7.6,-3.6)
   {children from the original $G-U$};
 \draw[impl] (p) -- (h);
 \node[lab,anchor=east] at (-0.3,-6.8) {node edges};
 \draw[impl] (d.south) -- (h.north east);
 \node[lab] at (4.35,-8.2) {child edges};
\end{tikzpicture}
\caption{Construction at one recursion node. Thick arrows pass constructed objects
to the next routine; the dashed arrow supplies a correctness condition.
The children are formed from the original node graph. The matrix data structure
of Section~\ref{sec:algebra} implements the queries and
restrictions inside these routines.}
\label{fig:architecture}
\end{figure}
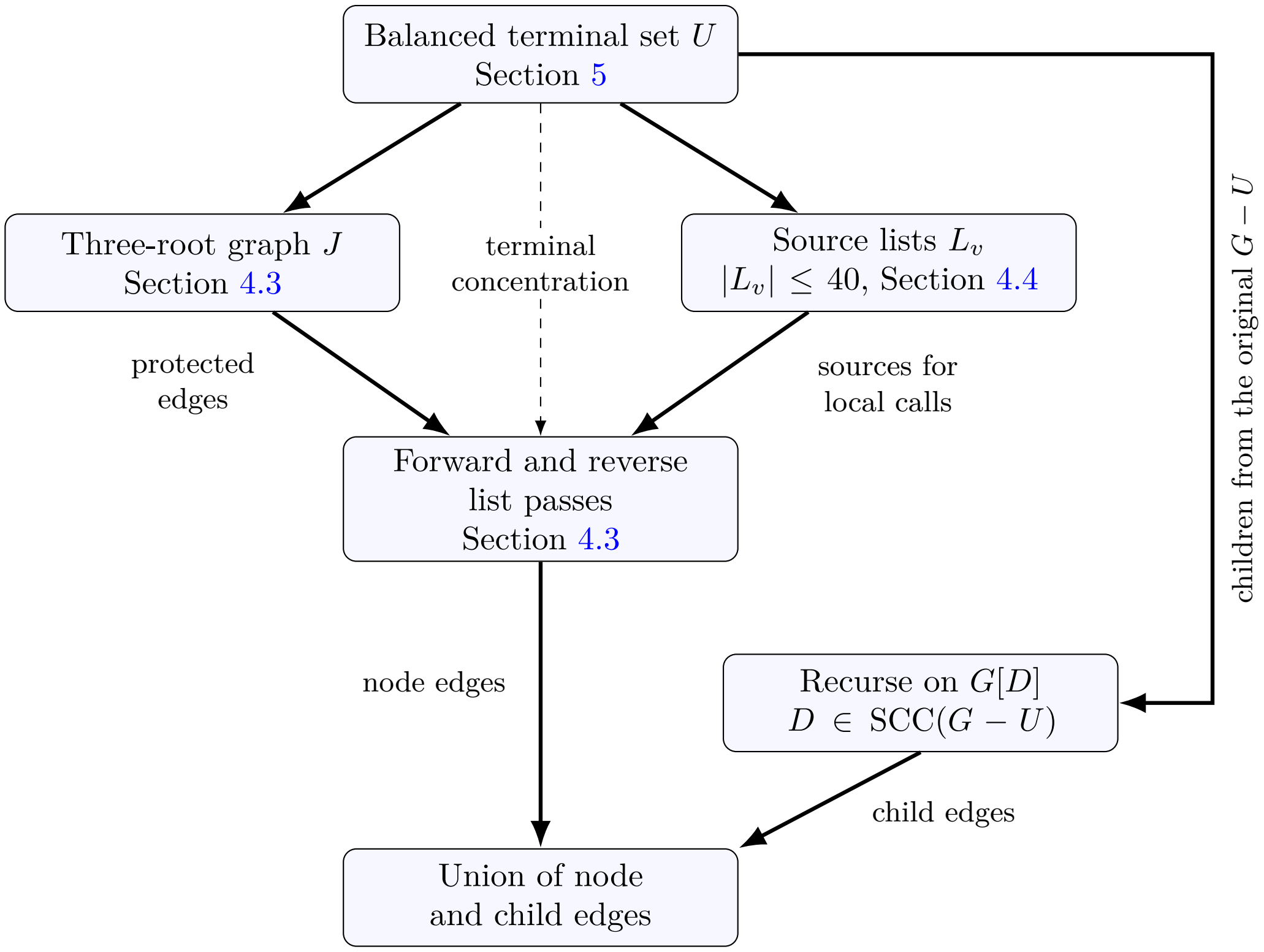

The lower bound and the construction are independent.
Section~\ref{sec:overview-lower} describes the lower-bound witness.
For the construction, Sections~\ref{sec:overview-recursion}--\ref{sec:overview-lists}
explain how to preserve the required SCCs once a suitable terminal set
has been chosen. Section~\ref{sec:overview-terminals} constructs that set,
and Section~\ref{sec:overview-time} implements the local operations and
derives the running time. Figure~\ref{fig:architecture} summarizes the
construction at one recursion node.

First decompose the input into SCCs and process their induced graphs
separately. Edges between them can be discarded: deletions cannot merge
SCCs, and paths connecting vertices within an original SCC stay inside
it. A one-vertex SCC needs no edges.

Throughout the algorithmic overview, $G$ is the strongly connected graph
at one recursive call, $n$ is its number of vertices, and $N$ is the
number of vertices in the full input. A failure set $F\subseteq E(G)$
contains at most two edges; a \emph{failure SCC} is an SCC of $G-F$.
For a current subgraph $X\subseteq G$, $X-F$ deletes the edges of $F$
still present in $X$. The notation $G-U$ deletes a vertex set $U$ and
all incident edges. The graph $G^{\rev}$ reverses every edge; failure
sets are reversed as well when working in this orientation.
Running-time bounds use a fixed achievable matrix-multiplication exponent
$2<\omega'<3$ over the algorithm's finite field, and $\softO$ suppresses
factors polynomial in $\log(N+2)$.

\subsection{Two failed path edges expose a necessary return edge}\label{sec:overview-lower}

Number $n=2^d$ vertices from zero to $n-1$. Include every forward path
edge $i\to i+1$. Also include $x\to x-2^j$ whenever bit $j$ of $x$ is
one: this operation clears that bit and leaves all other bits unchanged.
There are $n/2$ backward edges at each of the $d$ bit positions.

\begin{figure}[htbp]
\centering
\begin{tikzpicture}[>=stealth,scale=1.15,
  vertex/.style={circle,draw,minimum size=6mm,inner sep=1pt}]
\foreach \i in {0,...,7} {\node[vertex] (v\i) at (\i,0) {$\i$};}
\foreach \i in {0,4,5} {\node[vertex,fill=blue!12] at (v\i) {$\i$};}
\foreach \i/\j in {1/2,2/3,3/4,4/5,6/7}
  {\draw[->] (v\i) -- (v\j);}
\draw[->,red!75!black,dashed] (v0) -- node[above=3pt]{\scriptsize failed} (v1);
\draw[->,red!75!black,dashed] (v5) -- node[above=3pt]{\scriptsize failed} (v6);
\draw[->,blue!70!black,thick] (v5) to[bend left=30]
  node[below=3pt] {$e=(5,1)$} (v1);
\end{tikzpicture}
\caption{The witness for $5\to1$ in $G_3$. The shaded vertices form
$W=\{0,4,5\}$. Its full outgoing cut is
$\{0\to1,5\to6,5\to1\}$. Only the forward path and the selected backward
edge are drawn.}
\label{fig:witness}
\end{figure}
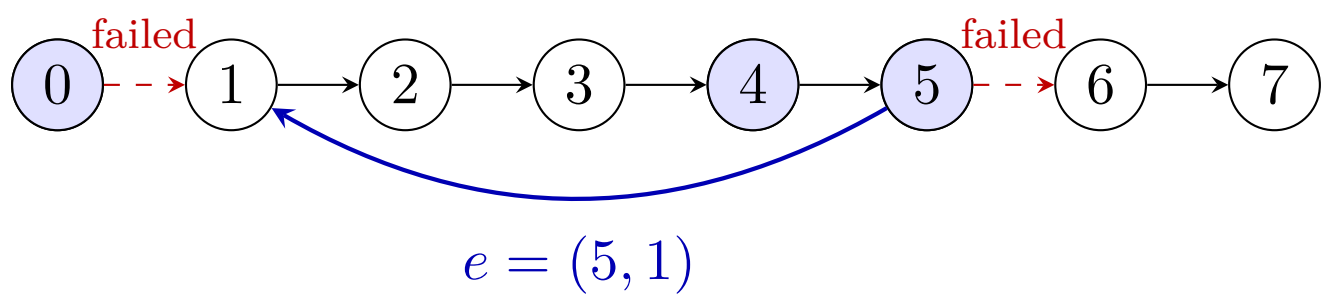

Fix a backward edge $e=b\to a$. Let $F_e$ contain the forward path edge
just before $a$ and the one just after $b$, when they exist. The forward
$a$--$b$ path and $e$ survive these at most two failures, so $a$ and $b$
remain in the same SCC. Section~\ref{sec:lower},
Lemma~\ref{lem:mandatory-backward}, proves that there is a vertex set
$W$ containing $b$ and excluding $a$ whose edges to the complement
are exactly $F_e\cup\{e\}$. Consequently, deleting $e$ as well leaves no path from
$b$ to $a$ and splits their SCC. Every valid preserver must therefore
retain $e$. Applying this witness to all backward edges gives the
$\Omega(n\log n)$ lower bound. Figure~\ref{fig:witness} illustrates the cut.

\subsection{The recursive task: cover terminals, then recurse}\label{sec:overview-recursion}

At each recursion node, choose a nonempty vertex set $U\subseteq V(G)$,
whose vertices we call \emph{terminals}. Preserve every failure SCC
that intersects $U$, and recurse on each induced subgraph $G[D]$ for
$D\in\SCC(G-U)$. A failure SCC avoiding $U$ remains strongly connected
in $G-U$, since its connecting paths lie inside that SCC.
It therefore lies in a single child $D$.

We require two properties of $U$:
\begin{description}[leftmargin=0pt,labelindent=0pt]
\item[Terminal concentration.] For every $F$, an SCC of $G-F$ contains
all but at most two terminals.
\item[Balance.] Every SCC of $G-U$ has at most $n/2$ vertices.
\end{description}
The terminal concentration property limits the work needed to preserve
SCCs meeting $U$. Balance bounds each child's size and hence the depth
and cost of recursion. The whole graph $G-U$ can still have $n-1$
vertices, which is why we recurse on its individual SCCs.
An SCC containing all but at most two terminals need not contain more than half
the vertices. We reserve \emph{vertex-large} for an SCC with more than
$n/2$ vertices.

A \emph{rooted preserver} for a root $u$ is a spanning subgraph of $G$ that
preserves reachability from $u$ to every vertex after every failure set
$F$. BCR construct such a graph with $O(n)$
edges~\cite{bcr}. Applying the construction to $G^{\rev}$ and reversing
its output edges back gives a graph preserving reachability to $u$.
Taking both graphs for every $u\in U$ would preserve each failure SCC
meeting $U$: any terminal in that SCC connects all its vertices in both
directions. This gives only an $O(n|U|)$ edge bound, however, and $U$
may have $\Theta(n)$ vertices. Our goal is to contribute $O(n)$ edges
at each recursion node. Section~\ref{sec:overview-local} introduces the
operations used to achieve this bound.

\Needspace{8\baselineskip}
\subsection{Local calls and rooted passes}\label{sec:overview-local}

An \emph{incoming restriction} at target $v$ retains a chosen subset
of its incoming edges and deletes the others. All edges with a different
head remain. In the current graph $X$, a \emph{local call}
$\Local(X,u,v)$, with $u\ne v$, selects at most six incoming edges
of $v$. Restricting $v$ to those edges preserves the $u$--$v$
reachability present in $X-F$, for every $F$. The call itself leaves
$X$ unchanged; the selected edges provide this guarantee together with
the rest of $X$. Section~\ref{sec:local} derives this guarantee from
BCR's farthest-cut selection argument~\cite[Lemma 13]{bcr}.

A \emph{rooted pass} fixes $u$ and processes every other target in turn,
making a local call in the current graph and then restricting that target.
BCR's locality lemma~\cite[Lemma 7]{bcr} explains why all required
reachability survives. Fix $F$ and a simple $u$--$w$ path in the current
$X-F$.
If a restriction at $v$ affects the path, the local call supplies a
replacement $u$--$v$ prefix after the restriction. The old $v$--$w$
suffix survives because it never re-enters $v$. Their concatenation
contains a $u$--$w$ path. Repeating this argument after each restriction
proves the rooted guarantee. Incoming edges of $u$ can also be discarded,
since a simple path starting at $u$ never uses them. The rooted pass
therefore produces a rooted preserver with at most $6n$ edges.

To preserve reachability from several sources to one target $v$, make a separate local
call for each source, all in the same current graph $X$. After collecting
their choices, retain their union and perform one incoming restriction
at $v$. For each source, this keeps every edge that its own call would retain, so
its guarantee still holds. Section~\ref{sec:overview-cover} uses this
observation to assign a short source list to each target.

\subsection{Three roots and source lists preserve SCCs meeting \texorpdfstring{$U$}{U}}\label{sec:overview-cover}

Choose any three terminals as roots, or all terminals if there are fewer.
Let the \emph{three-root graph} $J$ be the union of their forward and
reverse rooted preservers, with reverse edges mapped back to $G$. Then $|E(J)|\leq36n$.
Fix $F$ and a failure SCC $C$ containing more than two terminals.
By the terminal concentration property, some SCC of $G-F$ contains all
but at most two terminals. Since $C$ itself contains more than two,
it must be that SCC.
Thus $|U\setminus C|\leq2$, and the selected roots intersect $C$.
For a root in this intersection, its two rooted preservers keep $C$
strongly connected.

To handle the remaining failure SCCs meeting $U$, define $T_v$ to be
the set of terminals $u\ne v$ that share an SCC with $v$ after some
failure set $F$, where that SCC contains at most two terminals.
We construct a \emph{source list} $L_v$ for each target $v$ satisfying
\[
 T_v\subseteq L_v\subseteq U\setminus\{v\},
 \qquad |L_v|\leq40.
\]
Section~\ref{sec:overview-lists} explains the construction, and
Section~\ref{sec:lists} proves its guarantees. These lists are computed
from the original node graph and remain fixed during the passes.

The forward \emph{list pass} starts from $G$ and processes each target
$v$ once. At target $v$, it makes a local call for every source $u\in L_v$,
all in the same current graph. It then retains the union of their choices
and the incoming edges of $J$ at $v$, and performs one incoming restriction
at $v$. An independent list pass starts from $G^{\rev}$, uses the same
lists, and protects $J^{\rev}$. Each local call selects at most six
edges, so the number of edges in the union of the two pass outputs is at most
\[
 |E(J)|+12\sum_v|L_v|\leq516n.
\]

\Needspace{12\baselineskip}
To see why the lists suffice, fix a failure SCC $C$ with
$1\leq|C\cap U|\leq2$ and choose $u\in C\cap U$. We maintain that
$u$ reaches every $w\in C$ in the current graph after deleting $F$. This holds
at the start of the forward pass. Every surviving $u$--$w$ path lies
inside $C$: a vertex on it can return to $u$ through $w$ in the original
$G-F$, so it belongs to $C$, the SCC of $G-F$ containing $u$. A target $v$ whose
restriction could affect a simple such path therefore lies in
$C\setminus\{u\}$. The pair $C,F$ witnesses $u\in T_v\subseteq L_v$,
so the required local call at target $v$ supplies a surviving prefix;
the old suffix remains intact, as in Section~\ref{sec:overview-local}.
The reverse pass preserves paths from every $w\in C$ back to $u$.
Their union thus preserves $C$.

SCCs with more than two terminals are preserved by $J$, those with one
or two by the list passes, and those avoiding $U$ by the recursive
children. The output retains only original edges, so distinct failure
SCCs cannot be merged. Section~\ref{sec:construction}, Lemma~\ref{lem:preservation},
proves correctness of the full recursive construction by induction.

\subsection{Constructing the source lists with five anchors}\label{sec:overview-lists}

Our goal is to construct $L_v\supseteq T_v$ with $|L_v|\leq40$ for every
target $v$, without enumerating the failure sets defining $T_v$.
If $|U|\leq4$, take $L_v=U\setminus\{v\}$. Otherwise fix five terminals,
called anchors. We use them to find small lists that contain every
required source, following HSW's candidate-source and cut-container
approach~\cite[Section 4.2 and Lemma 4.1]{hsw}.

\Needspace{6\baselineskip}
For the correctness argument, consider $u\in T_v$ and a witnessing
SCC $C$ of $G-F$. The terminal concentration property supplies an SCC
$K_F$ containing all but at most two terminals. Since $|U|\geq5$,
$K_F$ contains at least three terminals, whereas $C$ contains at most
two; hence $C\ne K_F$. In at least one orientation, $v$ cannot reach
$K_F$: reachability in both orientations would make these two SCCs
mutually reachable in $G-F$.

\Needspace{4\baselineskip}
Write $\widehat G$ for this orientation and $\widehat F$ for the
correspondingly oriented failed edges. Let $W$ be the vertices reachable
from $v$ in $\widehat G-\widehat F$. Then $u\in W$ and $W$ avoids $K_F$,
so $|W\cap U|\leq2$. Every original edge leaving $W$ belongs to
$\widehat F$, since a surviving outgoing edge would reach another
vertex. Also, each vertex of $W$ has a path from $v$ staying inside
$W$. At least three of the five anchors therefore lie outside $W$;
let $Y$ be such a triple.

For fixed $v$, $Y$, and orientation, the two-failure case of HSW's
cut-container method either certifies that no such $W$ exists or returns
a set $S$ that contains every such $W$, excludes $Y$, and has at most
two outgoing original edges. We call $S$ a \emph{container}.
The computation uses $\widehat G,v,Y$ and does not require the unknown
$F$ or $W$.
Section~\ref{sec:lists}, Lemma~\ref{lem:container}, proves the particular
implementation used here.

To bound the number of sources recorded from a container, we must bound
$|S\cap U|$ as well. Let $F_S=\delta^+_{\widehat G}(S)$ be the original
edges leaving $S$. This is a failure set of size at most two.
After deleting $F_S$, no edge leaves $S$, so no SCC can intersect both
$S$ and its complement. Apply the terminal concentration property to
this new failure set; reversal does not change SCCs. The resulting SCC
$K_{F_S}$ cannot lie inside $S$, because all three terminals of $Y$
would then lie outside $K_{F_S}$, contradicting terminal concentration.
It is therefore disjoint from $S$, which gives
$|S\cap U|\leq2$.

The algorithm tries both orientations and all ten anchor triples,
skipping triples containing $v$. It records the terminals other than
$v$ in every returned container and takes their union as $L_v$.
There are at most twenty containers, each contributing at most two
terminals. For each $u\in T_v$, the preceding argument identifies a
triple whose container contains $u$, so the algorithm records it.
Thus $T_v\subseteq L_v$ and $|L_v|\leq40$ hold simultaneously for all
failure sets. Section~\ref{sec:lists} constructs all lists in
$\softO(n^{\omega'})$ time by sharing static matrix inverses across
vertices, using the query interface of Section~\ref{sec:algebra}.

\subsection{Constructing \texorpdfstring{$U$}{U} while keeping the graph fixed}\label{sec:overview-terminals}

We now construct a nonempty terminal set satisfying both terminal
concentration and balance. The SCCs of $G-F$ depend only on $G$ and $F$;
changing $U$ changes only how many terminals they contain. We exploit
this separation by representing the relevant SCCs once and maintaining
their terminal counts. The operation that changes $U$ follows HSW's
terminal exchange~\cite[Appendix C.2]{hsw}. Here we enforce
the terminal concentration property from Section~\ref{sec:overview-recursion}
and use a data structure to find the required exchanges efficiently.

\paragraph{Choosing which terminals to keep.}
There are two cases. If some failure set $F$ leaves no vertex-large SCC,
take $U$ to be the heads of its failed edges. Then $|U|\leq2$, so terminal
concentration holds automatically. Deleting $U$ removes every edge of
$F$, and hence every SCC of $G-U$ lies in an SCC of $G-F$ of size at most
$n/2$. This proves balance. Also $U$ is nonempty, since $G$ is strongly
connected and therefore this $F$ is nonempty.

Otherwise, every failure set has a unique vertex-large SCC. Initialize
$U=V(G)$, so initially $|U|=n$ and balance holds. If a failure SCC $C$ is vertex-large
and has at least three terminals outside it, let $F$ be its failure set
and replace $U$ by
\[
 U'=(U\cap C)\cup\{\operatorname{head}(e):e\in F\}.
\]
Deleting the added endpoints removes every edge of $F$, so an SCC of
$G-U'$ lies inside an SCC of $G-F$. Outside $C$ there are fewer than
$n/2$ vertices altogether. Inside $C$, every old terminal remains
deleted, so each new SCC lies in an old SCC of $G-U$. Thus balance
persists. A failure set requiring an exchange is nonempty, since
$F=\varnothing$ has $C=V(G)$; its endpoints keep $U'$ nonempty.
Each terminal exchange decreases $|U|$ and inserts at most two terminals.
There are therefore $O(n)$ exchanges and $O(n)$ individual insertions
and deletions. When no such $C$ remains, terminal concentration holds
for every failure set. Section~\ref{sec:exchanges}
proves these invariants and bounds.

\paragraph{Representing vertex-large failure SCCs.}
We first reduce the number of failure sets that need to be considered.
GIP's one-failure SCC certificate is a spanning subgraph $K\subseteq G$
with $O(n)$ edges that preserves every SCC partition after at most one
edge failure~\cite[Theorem 9.1]{gip}. Failure sets contained in $E(K)$
give $O(n^2)$ representatives, including sets of size zero or one.
For any other $F$, put $A=F\cap E(K)$. Then $|A|\leq1$ and
\[
 K-A\subseteq G-F\subseteq G-A.
\]
The outer graphs have the same SCC partition by the certificate
guarantee, so the middle graph has that partition as well. Thus the
representative $A$ accounts for $F$.

Next sample $O(\log N)$ vertices independently and uniformly from the
fixed $V(G)$, and call them \emph{probes}. A vertex-large SCC occupies
more than half of $V(G)$, so $t$ probes miss it with probability less
than $2^{-t}$. A union bound over the representatives shows that every
vertex-large failure SCC contains a probe with high probability.
For each probe $r$, let $Q_r$ be the union of its forward and reverse
rooted preservers, with edges mapped back to $G$. It has at most $12n$
edges. After every $F$, the SCC containing $r$ in $Q_r-F$ equals the
one containing $r$ in $G-F$, because the two rooted preservers retain
reachability in both directions through $r$. These rooted passes take
only $G$ and $r$ as input; they are performed before $U$ is chosen.

Allow $f,g\in E(K)\cup\{\bot\}$, where $\bot$ means no deletion.
Fix a probe $r$ and a first failure $f$, and let $B$ be the vertex set of
the SCC containing $r$ in $Q_r-f$. If $|B|\leq n/2$, discard this pair:
a further deletion cannot produce a vertex-large root component.
Otherwise work in $H=(Q_r-f)[B]$, which has $O(n)$ edges. The second
failure $g$ is now a one-failure problem in $H$.
A dominator tree records which vertices every path
from its root to a given vertex must visit. GIP's theorem gives the SCC
containing $r$ after a second failure $g$ as
\[
 C_{r,f,g}=B\setminus(A_g\cup R_g).
\]
Here $A_g,R_g\subseteq B$ are the vertices that lose reachability from
and to $r$ in $H-g$, respectively. GIP represents each set by a subtree,
possibly empty, in the forward or reverse dominator tree of $H$ rooted at
$r$~\cite[Theorem 3.7]{gip}. A vertex remains in the root SCC exactly
when neither loss occurs.
For each fixed $r,f$, the trees are shared by all choices of $g$.
A \emph{component record} stores the failure set and the two subtree
roots. We retain records only for vertex-large SCCs. There are
$\softO(n)$ choices of $r,f$ and $O(n)$ choices of $g$ for each, so
$\softO(n^2)$ records suffice. Section~\ref{sec:terminals} constructs
these records in $\softO(n^2)$ time and space once the probe graphs
are available.

If a representative failure set $F$ has no record, compute $\SCC(G-F)$
exactly. If no SCC is vertex-large, return the small terminal set above;
otherwise report $\FAIL$, since a vertex-large SCC was missed. If every
representative has a record, initialize $U=V(G)$ and perform terminal
exchanges using the monitor.

\paragraph{Monitoring the terminal counts.}
For a recorded component $C=B\setminus(A_g\cup R_g)$, maintain
\[
 |U\setminus C|
 =|U\setminus B|+|U\cap A_g|+|U\cap R_g|-|U\cap A_g\cap R_g|.
\]
\Needspace{8\baselineskip}
Only membership in $U$ changes; the graph, components, and trees stay
fixed. The implementation in Section~\ref{sec:terminals} and
Appendix~\ref{sec:monitor} has the following guarantee.
\begin{quote}
\textbf{Terminal monitor (Lemma~\ref{lem:monitor}).}
After $\softO(n^2)$ preprocessing, the component records occupy
$\softO(n^2)$ space and support one vertex insertion into or deletion from $U$
in $\softO(n^{3/2})$ time. In $\softO(n)$ time, the monitor finds a
record with at least three terminals outside its component, or
establishes that no such record exists.
\end{quote}
The proof converts changes of terminal membership into range updates
in the fixed dominator trees.
Once the monitor identifies a record, recover its component by an SCC
search in the corresponding $O(n)$-edge graph $Q_r-F$, and perform the
exchange through individual vertex insertions and deletions. Over all
exchanges, the $O(n)$ membership changes cost $\softO(n^{5/2})$
in total; finding and recovering the components costs another $\softO(n^2)$.
The remaining major cost is the $O(\log N)$ rooted passes for the probes,
bounded in Section~\ref{sec:overview-time}. The probe graphs are auxiliary
graphs used to choose $U$; their edges are discarded before constructing
$J$ and running the two list passes.

\subsection{The matrix data structure and the running time}\label{sec:overview-time}

In this section, we explain how the matrix data structure implements
rooted passes and list passes efficiently. To implement a local call,
we compute flow values up to a constant threshold, identify the relevant
minimum-cut side, and recover the original edges on its boundary.
CKS encode degree-constrained edge subsets by determinants and use this
encoding for bounded-flow queries~\cite[Sections 3.1 and 4.1]{cks}.
Section~\ref{sec:algebra} adapts their reduction to the queries used here.
The degree constraints are constant, so each graph vertex contributes
a constant-size block of rows and columns, giving matrix dimension
$O(n)$. The same section gives procedures for recovering original edge
identifiers. Together, these reductions implement one local call using
polylogarithmically many full inverse rows or columns and $\softO(n)$ additional work
(Proposition~\ref{prop:interface}). The data structure maintains the
determinant and access to rows and columns of the current inverse as
incoming restrictions change the graph.

The same data structure serves rooted passes and list passes. A rooted
pass retains at most six incoming edges per target. A list pass retains
at most six per source in $L_v$, together with the incoming edges of $J$.
Thus, writing $d_v$ for the retained indegree at target $v$, both kinds
of pass satisfy $\sum_v d_v=O(n)$. The protected edges of $J$ count
toward this total. At each target, the pass makes only a constant number
of local calls before its single incoming restriction. These are the conditions
needed by the following statement.
\begin{quote}
\textbf{Matrix pass (Lemma~\ref{lem:batch}).}
Consider a pass using the query interface of
Proposition~\ref{prop:interface}. Each target is restricted once.
At target $v$, the pass requests polylogarithmically many full inverse
rows or columns and retains $d_v$ incoming edges, with $\sum_vd_v=O(n)$. If the query answers
are correct and the matrices maintained for the current graph are
nonsingular, the pass takes
\[
 \softO\left(\frac{n^{\omega'+1}}{b}+n^2b+n^2\right)
\]
time and $\softO(n^2)$ space, for any integer $1\leq b\leq n$.
\end{quote}

We use standard low-rank inverse updates and periodic
rebuilding~\cite[Section 2]{bns}. An \emph{epoch} processes at most $b$
targets between rebuilds. At its start, compute the current encoding
matrix $M_0$, its inverse, and its determinant. Keep these fixed while
recording subsequent restrictions separately. Each restriction changes
a constant-size column block, so at most $O(b)$ columns differ from
$M_0$ during the epoch. The maintained correction supplies a full row
or column of the current inverse in $O(nb)$ time. At the next epoch,
rebuild from the graph with all preceding restrictions applied.

There are $O(n/b)$ rebuilds, each costing $\softO(n^{\omega'})$, giving
$\softO(n^{\omega'+1}/b)$ time. The $\softO(n)$ requests for full inverse
rows or columns cost $\softO(n^2b)$ time. Larger epochs reduce
the rebuilding cost while increasing the cost of answering these requests.
Maintaining the correction data fits within the $\softO(n^2b)$ bound.
The $\softO(n^2)$ term covers forming the updates, using
$\sum_v d_v=O(n)$, and the remaining local computations.
Section~\ref{sec:batch} proves these implementation bounds.

Taking
\[
 b=\left\lceil n^{(\omega'-1)/2}\right\rceil,
 \qquad a=\frac{\omega'+3}{2},
\]
balances the costs of rebuilding and answering inverse queries, giving
$\softO(n^a)$ time per pass.
At one recursion node, the algorithm makes $O(\log N)$ rooted passes
for probes, at most six rooted passes for the three selected terminals,
and two list passes. Their combined cost is $\softO(n^a)$.
The other main costs are $\softO(n^{\omega'})$ for source lists and
$\softO(n^{5/2})$ for the terminal monitor. Both are smaller because
$\omega'<a$ and $5/2<a$ when $2<\omega'<3$.

All work at a node before its recursive calls is therefore $\softO(n^a)$.
At each successive depth, balance and disjoint child vertex sets
reduce the sum of the $n^a$ bounds by a constant factor.
Thus the total time over the recursion forest is $\softO(N^a)$;
Section~\ref{sec:global} gives the summation.
At each depth the node graphs have at most $N$ vertices in total, so
their $O(n)$ edge contributions sum to $O(N)$. The $O(\log N)$ recursion
depth gives $O(N\log N)$ output edges. Processing children one at a time
uses $\softO(N^2)$ space.

The principal random choices are the vertex probes and the finite-field
evaluations of the CKS determinant encoding. Field construction and
uniform sampling also use random bits. Undetected evaluation errors
make the construction Monte Carlo. Section~\ref{sec:probability} and
Appendix~\ref{sec:field} choose the sample counts and field sizes so that,
for any fixed $c>0$, the algorithm returns a correct preserver within
the stated bounds with probability at least $1-N^{-c}$. They also
account for adaptive queries and the bit cost of field arithmetic.

\section{The size lower bound}\label{sec:lower}

The lower bound is independent of the construction algorithm. Its graph
contains a forward path and backward edges at several binary scales.
For each backward edge, two failed path edges block all detours that
could replace it. Its endpoints remain in the same SCC after those
failures, but deleting the backward edge separates them.

\subsection{The graph and its size}

Fix $d\geq1$ and put $n=2^d$. Let $G_d$ have vertex set
$\{0,1,\ldots,n-1\}$. Include the following edges:
\begin{enumerate}
\item The forward path edges $(i,i+1)$ for $0\leq i<n-1$.
\item For each vertex $x$ and each set bit $j$ of $x$, the backward edge
      $(x,x-2^j)$.
\end{enumerate}
Bit positions start at zero. The second rule clears a set bit. It does not
include a backward jump for a bit that is zero.

For each of the $d$ bit positions, exactly $n/2$ vertices have that bit set.
All backward edges are distinct. They are also distinct from the forward
edges. Therefore
\begin{equation}\label{eq:lower-count}
 |E(G_d)|=n-1+\frac{dn}{2}
         =n-1+\frac n2\log_2n.
\end{equation}
The graph is simple and strongly connected. A vertex reaches zero by
clearing its set bits. Zero reaches every vertex along the forward path.

\subsection{A failure witness for every backward edge}

\begin{lemma}\label{lem:mandatory-backward}
Every backward edge of $G_d$ belongs to every two-edge-failure SCC preserver
of $G_d$.
\end{lemma}

\begin{proof}
Fix a backward edge $e=(b,a)$ with $b=a+2^j$. Bit $j$ of $a$ is zero.
Fail the forward path edges immediately outside the interval $[a,b]$:
\begin{equation}\label{eq:lower-failures}
 F_e=\{(a-1,a):a>0\}\ \cup\ \{(b,b+1):b<n-1\}.
\end{equation}
A conditional set in this expression is empty when its condition is false.
Thus $|F_e|\leq2$, and $e\notin F_e$. The cycle
\[
 a\longrightarrow a+1\longrightarrow\cdots\longrightarrow b
   \longrightarrow a
\]
survives in $G_d-F_e$. In particular, $a$ and $b$ are in the same SCC.

We must show that every surviving path from $b$ to $a$ uses $e$.
A path cannot move above $b$, since the forward edge leaving $b$ has
failed. If it moves below $a$, it cannot return across the other failed
path edge. To handle the remaining backward jumps, split the binary
block containing $a$ and $b$ into halves. Let $c$ be the start of its
upper half; $a$ and $b$ have the same offset $r$ in their respective
halves. Thus
\[
 a=z2^{j+1}+r,\qquad 0\leq r<2^j,\qquad
 c=z2^{j+1}+2^j,\qquad b=c+r.
\]
The interval $[c,b]$ contains $b$. Add the prefix below $a$ so that
backward jumps below $a$ also stay in the proposed cut side. Define
\begin{equation}\label{eq:lower-cut}
 W=\{0,\ldots,a-1\}\ \cup\ \{c,\ldots,b\}.
\end{equation}
The first interval is empty when $a=0$. Vertex $b$ is in $W$, and $a$ is
outside $W$. We claim that
\begin{equation}\label{eq:exact-cut}
 \delta_{G_d}^+(W)=F_e\cup\{e\}.
\end{equation}

The forward edges leaving $W$ are exactly the edges in $F_e$.
A backward edge from a vertex below $a$ stays below $a$.
Consider a backward edge from $y\in\{c,\ldots,b\}$ that clears bit $h$.
There are three cases.
\begin{enumerate}
\item If $h<j$, clearing it keeps the endpoint at least $c$ and at most $b$.
      The edge stays in $W$.
\item If $h=j$, the endpoint is at most $a$. It equals $a$ only when
      $y=b$. This is the edge $e$. Every smaller endpoint belongs to $W$.
\item If $h>j$, the endpoint is below $a$, since
      $y-2^h\leq b-2^{j+1}<a$. The edge stays in $W$.
\end{enumerate}
This proves~\eqref{eq:exact-cut}.

After deleting $F_e$ and $e$, no edge leaves $W$. Hence $b$ cannot reach
$a$ in $(G_d-e)-F_e$. Their SCC has split. Any subgraph that omits $e$
also fails this requirement. Thus every valid preserver retains $e$.
\end{proof}

\subsection{Forward edges and arbitrary vertex counts}

\begin{lemma}\label{lem:mandatory-forward}
Every forward edge of $G_d$ is necessary even without failures.
\end{lemma}

\begin{proof}
For $0\leq i<n-1$, the prefix $P_i=\{0,\ldots,i\}$ has exactly one
outgoing edge, namely $(i,i+1)$. All backward edges from $P_i$ stay in it.
Deleting $(i,i+1)$ destroys strong connectivity. Since $G_d$ is strongly
connected, every zero-failure SCC preserver must retain this edge.
\end{proof}

Lemmas~\ref{lem:mandatory-backward} and~\ref{lem:mandatory-forward} show
that every edge in~\eqref{eq:lower-count} is necessary. Thus $G_d$ is its
own unique two-failure preserver.

To obtain every sufficiently large vertex count, we add vertices while
keeping the old backward-edge witnesses valid. For $N\geq4$, let
$q=2^{\lfloor\log_2N\rfloor}$. Start with $G_{\log_2q}$ and attach a bidirected path from $q-1$ through
$N-q$ new vertices. The new graph remains simple and strongly connected.
For an old backward edge $(b,a)$, use~\eqref{eq:lower-failures}, with
$(b,b+1)$ included whenever it exists in the extended path.
Keep the same set $W$ from~\eqref{eq:lower-cut}. A new edge can leave $W$
only when $b=q-1$, and then it is the failed edge $(b,b+1)$.
Thus~\eqref{eq:exact-cut} and the surviving cycle still apply.
At least $(q/2)\log_2q=\Omega(N\log N)$ edges remain necessary.

\subsection{The same graphs have linear one-failure preservers}

Deleting a forward path edge leaves a prefix and a suffix. We choose
two families of backward edges so that every prefix and every suffix
is strongly connected. These families also provide two edge-disjoint
return paths when a backward edge fails.
Let $K_d$ contain the full forward path and the following edges:
\begin{enumerate}
\item For each $x>0$, include the edge that clears the least significant
      set bit of $x$. Call this set $T_0$.
\item For each $x<n-1$, let $j$ be its least significant zero bit.
      Include $(x+2^j,x)$. Call this set $T_1$.
\end{enumerate}
Each set has $n-1$ edges. Therefore $|E(K_d)|\leq3(n-1)$.

\begin{lemma}\label{lem:linear-one-failure}
The subgraph $K_d$ preserves every SCC of $G_d$ after at most one edge failure.
\end{lemma}

\begin{proof}
The edges of $T_0$ give a path from $x$ to zero using only labels at most
$x$. Together with the forward path, they make every prefix strongly
connected. The edges of $T_1$ give a path from $n-1$ to $x$ using only
labels at least $x$. To obtain it, start at $x$, repeatedly set its least
significant zero bit, and reverse the resulting sequence. Thus every
suffix is also strongly connected.

If a forward edge $(i,i+1)$ fails, no remaining edge can go from
$\{0,\ldots,i\}$ to $\{i+1,\ldots,n-1\}$. Both induced subgraphs are
strongly connected in $K_d-(i,i+1)$ and hence in $G_d-(i,i+1)$.
They are exactly the two SCCs
in both graphs.

Suppose a backward edge fails and $d\geq2$. The forward path survives.
There are two edge-disjoint paths from $n-1$ to zero in $K_d$:
\[
\begin{split}
 n-1&\longrightarrow n-2\longrightarrow n-4\longrightarrow\cdots
       \longrightarrow n-2^{d-1}\longrightarrow0,\\
 n-1&\longrightarrow 2^{d-1}-1\longrightarrow2^{d-2}-1
       \longrightarrow\cdots\longrightarrow1\longrightarrow0.
\end{split}
\]
The first uses $T_0$, and the second uses $T_1$. Their internal vertices
are disjoint. One path survives the failure. It and the forward path
make all vertices strongly connected. This also proves the zero-failure
case. If $d=1$, then $K_d=G_d$, so the claim is immediate.
\end{proof}

This finishes the proof of Theorem~\ref{thm:lower-main}.
All failure sets $F_e$ for the backward-edge lower bound are contained in
the forward path, and hence in $K_d$. The obstruction remains even when
the two failures are restricted to a fixed linear-size one-failure preserver.

\subsection{Optimal size and the scope of the statement}

\begin{proof}[Proof of Corollary~\ref{cor:optimal-main}]
Theorem~\ref{thm:lower-main} gives $s_2(N)=\Omega(N\log N)$.
Every $k$-failure preserver for $k\geq2$ is also a two-failure preserver,
so $s_k(N)\geq s_2(N)$. The published upper bound
$s_k(N)=O(k4^kN\log N)$~\cite[Theorem 4.2]{hsw} proves
$s_k(N)=\Theta_k(N\log N)$ for every fixed $k\geq2$.

For $k=1$, a directed cycle requires all its $N$ edges even without
failures, giving $s_1(N)=\Omega(N)$. The linear-size single-failure
certificate of~\cite{gip} gives the matching upper bound.
\end{proof}

The logarithmic factor comes from $n/2$ necessary backward edges at
each of $\log_2n$ scales. The constants in $\Theta_k(N\log N)$ may
depend on $k$. The witness graphs contain strong bridges, that is,
edges whose deletion destroys strong connectivity; the lower bound
applies to subgraph preservers for general directed graphs.

\section{The recursive construction}\label{sec:recursive}

We first give the construction assuming that local calls answer correctly
and that a terminal set with the required properties is available.
The section proves preservation and the output-size bound, including the
construction of the source lists. Section~\ref{sec:terminals} constructs
the terminal set, and Section~\ref{sec:algebra} implements the local calls
and establishes the running time and success probability.

\subsection{The recursive task and terminal-set requirements}\label{sec:requirements}

At a recursive call on a strongly connected graph $G$, we preserve all
failure SCCs that meet a chosen nonempty terminal set $U$. We then
recurse on the induced graphs $G[D]$ for $D\in\SCC(G-U)$. The terminal
set must satisfy the two properties from the overview:
\begin{align}
 &\text{for all }F\subseteq E(G),\ |F|\leq2,\quad
 \max_{D\in\SCC(G-F)}|D\cap U|\geq |U|-2;
 &&\tag{G$_2$}\label{eq:giant}\\
 &\text{for all }D\in\SCC(G-U),\quad |D|\leq n/2.
 &&\tag{B}\label{eq:balance}
\end{align}
These are the terminal concentration and balance properties. Terminal concentration
holds automatically for $|U|\leq3$. When $|U|\geq5$, the SCC containing all but at most two terminals is
unique and contains at least three terminals. This component need not
have more than $n/2$ vertices. We use \emph{vertex-large} only for a
component with more than $n/2$ vertices.

For the initial input, compute its SCCs and process their induced graphs
separately. Deletions cannot merge these components, and every path
between vertices of one component stays inside it. The edges between
original SCCs can therefore be discarded. A one-vertex component
requires no edges.

\paragraph{Graph and failure conventions.}
We reserve $N$ for the number of vertices in the full input and $n$ for the number of vertices at
a recursion node. Vertices and original edges have fixed distinct
identifiers. Temporary query graphs may
have parallel edges or constant integer capacities; each unit copy has
its own identity. Auxiliary edges never appear in the output or in a failure set.
When unit copies encode an original edge, they are associated with the same
original edge identifier: deleting that edge removes all its copies.

For a graph $X$, $X-F$ deletes edges in $F\cap E(X)$, $X-U$ deletes
vertices in $U$, and $X[W]$ is the subgraph induced by $W$. The graph
$X^{\rev}$ reverses every edge and keeps its identity. The set
$\delta_X^+(W)$ consists of edges with tail in $W$ and head outside it.
For a capacitated graph, the capacity of a cut is the sum of those edge
capacities. A set is \emph{internally reachable from $v$} if it contains
$v$ and every one of its vertices is reachable from $v$ using only
vertices of the set. All paths can be taken simple unless stated otherwise.

\subsection{Local calls and rooted passes}\label{sec:local}

We select incoming edges using small flow values and minimum cuts.
We first fix the flow notation, then state the local call and explain
how sequential incoming restrictions produce a rooted preserver.

For distinct vertices $s,t$, $\flow_X(s,t)$ is the maximum integral flow
from $s$ to $t$. With unit capacities it equals the maximum number of
edge-disjoint paths. The max-flow/min-cut theorem~\cite{ahu} states that
\[
 \flow_X(s,t)=\min_{s\in C,\ t\notin C}
           \sum_{e\in\delta_X^+(C)}\operatorname{cap}(e).
\]
Here $C$ is a vertex set, its \emph{boundary} $\delta_X^+(C)$ is an edge
set, and $\operatorname{cap}(e)$ is the positive integer capacity of $e$.
The tails or heads of a boundary form vertex sets, with duplicates removed.
For unit capacities, cut submodularity states
$|\delta^+(A)|+|\delta^+(B)|\geq
 |\delta^+(A\cup B)|+|\delta^+(A\cap B)|$.
The \emph{flow value capped at $K$} is
$\min\{K,\flow_X(s,t)\}$; it is obtained by testing thresholds
$1,\ldots,K$ and counting the true thresholds. Our $K$ is always a
constant. SCCs are computed in linear time~\cite{tarjan}. The \emph{condensation} has one vertex per SCC and an edge for each edge
between distinct SCCs; it is acyclic.

An \emph{incoming restriction} at target $t$ permanently
deletes specified original edges with head $t$; all other edges are kept.
We distinguish this update from hypothetical failures $F$ in a proof and
from temporary query changes. In particular, tentative group deletions,
shortcuts, and capacity changes used to choose an incoming restriction
are undone after their query; only the chosen restriction changes $X$.

A local call selects the incoming edges to retain at one target.
Its guarantee will justify both rooted passes and list passes. It adapts BCR's farthest-cut selection~\cite[Lemmas 10 and 13]{bcr}
and uses their locality lemma~\cite[Lemma 7]{bcr} to obtain a full rooted
preserver. BCR obtain the stronger $4n$ bound for two failures; the
six-edge variant below uses at most four recursive invocations,
which makes its query requirements explicit.
Parallel unit edges are counted separately. A source side of an
$s$--$t$ cut contains $s$ and excludes $t$.

\begin{lemma}[Farthest-cut property; cf.~{\cite[Lemmas 9 and 10]{bcr}}]\label{lem:farthest}
Let $B$ be a directed graph with unit capacities, let $s\ne t$, and let its minimum $s$--$t$ cut value be $\lambda$.
There is a unique inclusion-wise maximum source side $C$ among all
minimum cuts. For $w\notin\{s,t\}$,
\[
 w\in C
 \quad\Longleftrightarrow\quad
 \flow_{B+(s,w)}(s,t)=\lambda,
\]
where $(s,w)$ is a new unit edge. If $w\notin C$, the new value is
$\lambda+1$.
\end{lemma}

The maximum side is the union of the minimum source sides, by cut
submodularity. Adding $s\to w$ raises every minimum cut exactly when
$w$ is outside this union. This is the farthest-cut property
of~\cite[Lemmas 9 and 10]{bcr}.

We call this unique maximum source side $C$ the \emph{farthest cut side}.

Fix a current graph $X$, a source $s$, and a target $t\ne s$.
The procedure $\Local(X,s,t,A)$ uses a set $A$ of at most two auxiliary unit edges, called \emph{shortcuts}. All edges in $A$ leave $s$. None enters $t$.
The output consists only of original edges of $X$ entering $t$.
The top-level call has $A=\varnothing$; we abbreviate it as
$\Local(X,s,t)$. The call returns an edge set and leaves $X$ unchanged.
Its caller applies the incoming restriction after collecting all local
choices for the target.

The local call operates on $B=X+A$ as follows.
\begin{enumerate}
\item If $\flow_B(s,t)=0$, return the empty set.
\item If $\flow_B(s,t)\geq3$, choose three incoming edges of $t$ such that
restricting $t$ to these edges leaves an $s$--$t$ flow of value at least three.
Return them.
\item Otherwise the value is $\lambda\in\{1,2\}$. Find the farthest cut side
$C$ and all edges of $\delta_B^+(C)$. For each such edge $(u,h)$:
\begin{itemize}
\item if $h=t$, retain this edge;
\item if $h\ne t$, add a fresh unit edge $(s,h)$, call the procedure
recursively, and retain its returned edges.
\end{itemize}
Return the union.
\end{enumerate}

\begin{lemma}\label{lem:local}
The top-level call returns at most six original incoming edges of $t$.
Restricting the incoming edges of $t$ to this set preserves $s$--$t$
reachability after every set of at most two original edge failures.
There are at most four recursive invocations, including the initial one.
\end{lemma}

\begin{proof}[Proof outline]
Each shortcut raises the flow by one. A call with flow value one has at most
one child with flow value two; a call with flow value two has at most two
children with flow value three. Each call that stops with flow at least
three returns three incoming edges.
For correctness, consider the first boundary edge on a surviving path.
A shortcut used by a recursive call can be replaced by the corresponding
prefix of the parent path, which never enters the restricted target. This is the selection
argument of BCR's Lemma~13~\cite{bcr}, with a preliminary shortcut when
the initial flow value is one.
Appendix~\ref{sec:primitive-details} gives the full argument for this
variant, including the distinction between original and auxiliary edges.
\end{proof}

\begin{corollary}[Locality principle of~{\cite[Lemma 7]{bcr}}]\label{cor:rooted}
For a fixed root $s$, sequential application of Lemma~\ref{lem:local} to
every target $t\ne s$ gives a rooted two-edge-failure reachability
preserver with at most $6n$ edges on an $n$-vertex graph.
\end{corollary}

\begin{proof}
When target $t$ is processed, take a current simple path from $s$ to any
vertex $w$ after a fixed failure set. If it enters $t$, replace its prefix
by the preserved $s$--$t$ path. Its suffix does not enter $t$ again.
Deleting cycles from the resulting walk gives a simple path for the
next restriction. Thus each restriction preserves all required reachability. A simple path
from $s$ does not use an incoming edge of $s$, so all these edges can be
removed. There are at most six retained edges at each other target.
\end{proof}

The local call is deterministic given exact flow and cut answers.
Proposition~\ref{prop:interface} implements it using
$\polylog(n+2)$ requests for full inverse rows or columns and
$\softO(n)$ additional work. We use the combinatorial guarantee above
until that implementation is established in Section~\ref{sec:algebra}.

\subsection{Three roots and source lists}\label{sec:node-passes}

Fix a nonempty terminal set $U$ satisfying terminal concentration.
We show how to preserve every failure SCC meeting $U$ using $O(n)$
edges. Balance will be used only when we recurse.

For each vertex $v$, define the required sources by
\begin{equation}\label{eq:true-list}
 T_v=\left\{u\in U\setminus\{v\}:\begin{array}{l}
 \text{for some }|F|\leq2,\text{ an SCC of }G-F\text{ contains }u,v\\
 \text{and contains at most two terminals}
 \end{array}\right\}.
\end{equation}
We seek source lists satisfying $T_v\subseteq L_v\subseteq U\setminus\{v\}$ and $|L_v|\leq40$.
Equivalently, for each failure SCC $C$ with one or two terminals and
each $u\in C\cap U$, every $v\in C\setminus\{u\}$ must list $u$.
The inclusion $T_v\subseteq L_v$ ensures that later restrictions preserve
the paths needed by earlier targets. The size bound limits the number
of local calls. Section~\ref{sec:lists} constructs these lists without
enumerating the failure sets that define $T_v$.

Choose the $\min\{3,|U|\}$ terminals with smallest identifiers as roots.
For each selected root, perform a rooted pass on a separate copy of
$G$ and another on a separate copy of $G^{\rev}$. Map reverse edges
back to their original orientation. Their union is the \emph{three-root
graph} $J$, with $|E(J)|\leq36n$.

The \emph{forward list pass} starts from $G$ and processes targets in
identifier order. At target $v$, make a local call $\Local(X,u,v)$ for
each $u\in L_v$, all in the same current graph $X$. Retain the union of
the selected edges and the incoming edges of $J$, then perform one
incoming restriction at $v$. If $L_v$ is empty, retain only these edges
of $J$. The \emph{reverse list pass} starts independently from
$G^{\rev}$, uses the same lists, and protects $J^{\rev}$ in the same
way. Map its output back and take the union of the two pass outputs.
Protected edges remain subject to hypothetical failures $F$; protection
means that the construction never deletes them.

\begin{lemma}[Preserving failure SCCs meeting $U$]\label{lem:node-preservation}
If $U$ satisfies terminal concentration, the source lists satisfy
$T_v\subseteq L_v$, and all local calls are correct, the union of the
two list-pass outputs preserves every failure SCC that meets $U$.
\end{lemma}
\begin{proof}
Fix $F\subseteq E(G)$ with $|F|\leq2$ and an SCC $C$ of $G-F$
meeting $U$.
Suppose first that $|C\cap U|>2$. If $|U|\leq3$, all terminals are
roots. Otherwise the terminal concentration property implies that $C$ leaves at
most two terminals outside itself. Indeed, the component supplied by
terminal concentration must be $C$, since all other components together have
at most two terminals. Thus the three selected roots cannot all miss
$C$. For a root in $C$, the two rooted preservers in $J$ preserve paths
from the root to each vertex of $C$ and back after $F$. This keeps $C$
strongly connected.

Next suppose $1\leq|C\cap U|\leq2$, and choose $u\in C\cap U$.
Let $X_i$ be the current graph after $i$ forward restrictions. We prove
the invariant that $u$ reaches every $w\in C$ in $X_i-F$. It holds for
$X_0=G$. Assume it for $X_{i-1}$, just before processing target $v$,
and choose a surviving simple $u$--$w$ path $P$ in $X_{i-1}-F$.
Every vertex of $P$ lies in $C$: it is
reachable from $u$ and can return to $u$ through $w$ in $G-F$.

If $P$ uses no incoming edge of $v$, it is unaffected; this includes
$v=u$. Otherwise $v\in C\setminus\{u\}$, so $u\in T_v\subseteq L_v$.
The call $\Local(X_{i-1},u,v)$ is therefore made before this restriction.
Its guarantee applies to the current graph, including all earlier
deletions. The combined retained set contains its entire choice, so
there is a surviving $u$--$v$ path $Q$ after the restriction. The old
suffix $P[v,w]$ never re-enters $v$ and remains intact. Thus
$Q\circ P[v,w]$ is a surviving walk; removing cycles gives a path.
This proves the invariant at the next step. It also shows why calls
for other sources at $v$ cause no conflict: their choices are unioned
before the single restriction.

At the end of the forward pass, $u$ reaches all of $C$. The independent
pass from $G^{\rev}$ proves reachability from all of $C$ back to $u$.
Their union keeps $C$ strongly connected.

\end{proof}

\begin{lemma}[Edges and calls at one node]\label{lem:node-size}
If $|L_v|\leq40$ for every target, the two list passes contribute at
most $516n$ edges. Each pass makes $O(n)$ local calls and satisfies
$\sum_v d_v=O(n)$, where $d_v$ counts all incoming edges retained at
target $v$, including protected edges. The at most six rooted passes
used for $J$ satisfy the same retained-indegree bound.
\end{lemma}
\begin{proof}
A list pass retains at most
$|E(J)|+6\sum_v|L_v|=O(n)$ edges and makes
$\sum_v|L_v|\leq40n$ local calls. The forward and reverse outputs
share $J$, so their union has at most
\[
 |E(J)|+12\sum_v|L_v|\leq36n+12\cdot40n=516n
\]
edges. A rooted pass retains at most six incoming edges per target.
\end{proof}

These bounds verify the conditions of the matrix-pass implementation in
Lemma~\ref{lem:batch}, using Proposition~\ref{prop:interface} to bound the
inverse queries per local call.

\subsection{Constructing the source lists}\label{sec:lists}

Throughout this subsection, $G$ is strongly connected and the nonempty
set $U\subseteq V(G)$ satisfies terminal concentration~\eqref{eq:giant}. Following the
candidate-source strategy of HSW~\cite[Section 4.2, Algorithm 1]{hsw},
we assign different sources to different targets. Their general
construction samples terminal subsets and uses their cut-container
lemma~\cite[Lemma 4.1]{hsw}. The two-failure specialization below
enumerates triples of five fixed terminals and uses the stronger
concentration property to obtain a constant list size.

Our goal is to construct the source lists required in
Section~\ref{sec:node-passes}: for every target $v$,
$T_v\subseteq L_v\subseteq U\setminus\{v\}$ and $|L_v|\leq40$.
The lists must cover the required sources for all failure sets
simultaneously. If $|U|\leq4$, take $L_v=U\setminus\{v\}$.
Otherwise choose the five smallest-identifier terminals as $R_5$.
This set is fixed for the list construction, separate from the sampled
vertex probes of Section~\ref{sec:terminals} and the selected terminal roots used for $J$.
For each of the $\binom53=10$ three-element sets $Y\subseteq R_5$,
work in both orientations.

For each such $Y$, add a sink $z$ and three parallel unit edges
$y\to z$ for each $y\in Y$; this is its \emph{anchor graph}. For each
$v\notin Y$, perform the following container construction. Strong connectivity and
$Y\ne\varnothing$ ensure positive flow from $v$ to $z$; an observed
zero causes $\FAIL$.
\begin{enumerate}
\item At flow at least three, return no set.
\item At flow two, return the farthest cut side $S$.
\item At flow one, let $S_0$ be the farthest cut side and $(a_0,h)$ its
unique boundary edge. Add a new unit edge $v\to h$. The flow becomes
two. Return its new farthest cut side $S$.
\end{enumerate}
The returned set is a \emph{container}: it contains every internally
reachable set specified in Lemma~\ref{lem:container} and excludes the anchors.
The construction specializes HSW's cut-container principle~\cite[Lemma 4.1 and
Section 6]{hsw} to cut value two. We prove the particular one-shortcut procedure because
its query count is used in the running-time bound.
The unique boundary edge in the last case is original. In particular,
$h\ne z$, because three parallel anchor edges cannot form a cut of
value one.

\begin{lemma}\label{lem:container}
Assume that $U$ satisfies terminal concentration~\eqref{eq:giant}
and $|U|\geq5$. Let $Y\subseteq U$ have size three, and let $v\notin Y$.
With exact flow answers, every set returned for $(v,Y)$ contains every set $W$ that contains $v$,
avoids $Y$, is internally reachable from $v$, and has at most two
outgoing original edges. If such a $W$ exists, the construction returns a container.
Each returned set has at most two terminals and at most two outgoing
original edges.
\end{lemma}

\begin{proof}
Such a $W$ is a cut side of value at most two in the anchor graph, so
the original flow is at most two. At value two, $W$ is a minimum source
side and lies in $S$.

At value one, if $W\subseteq S_0$, then $W$ lies in the new farthest
cut side because $S_0$ is a cut of value two after the shortcut is added.
Otherwise, an internal path in $W$ from $v$ to a vertex outside $S_0$
uses the unique exit $(a_0,h)$. Hence $h\in W$. The shortcut is internal
to $W$, which remains a cut of value at most two. It is a minimum cut
in the augmented graph and lies in the new farthest cut side.

Each returned $S$ excludes $Y$ and has at most two outgoing original
edges. Delete these edges from the original graph in the chosen
orientation. No SCC crosses $S$. The terminal concentration property supplies an
SCC with all but at most two terminals. It cannot lie in $S$, since
the three anchors lie outside $S$. Therefore it lies outside $S$, and
$S$ has at most two terminals.
\end{proof}

For each returned set, scan its membership vector and record its
terminals other than $v$. Let $L_v$ be the union of these recorded
identifiers, with duplicates removed. The target $v$ of a source list is the source of the auxiliary
$v$--$z$ flow query. If $v\in Y$, skip that anchor triple.

\begin{proposition}\label{prop:lists}
Assume all threshold answers are correct and the maintained matrices
are nonsingular. The procedure returns lists such that, for every $v$,
\[
 T_v\subseteq L_v\subseteq U\setminus\{v\},\qquad |L_v|\leq20\cdot2=40.
\]
It computes all lists in $\softO(n^{\omega'})$ time and $\softO(n^2)$ space.
Detected inconsistencies cause $\FAIL$ as in Proposition~\ref{prop:interface}.
\end{proposition}

\begin{proof}
If $|U|\leq4$, both claims about the lists are immediate. Assume
$|U|\geq5$. There are $2\binom53=20$ choices of orientation and anchor
set, and each returns at most two terminals. This proves the bound
$20\cdot2=40$ by Lemma~\ref{lem:container}. Consider
$u\in T_v$ and its witnessing SCC after failures $F$. Since $|U|\geq5$,
the SCC supplied by the terminal concentration property has at least three terminals
and is different from the witness. In at least one orientation, $v$
cannot reach this component: reachability to it in both orientations
would make $v$ mutually reachable with it in $G-F$, contradicting that
the SCCs differ. Let $W$ be all vertices reachable
from $v$ in that orientation of $G-F$. This set is internally reachable,
contains $u$, has outgoing original boundary contained in $F$, and has
at most two terminals. At least three of the five vertices of $R_5$ lie
outside $W$. The corresponding choice of $Y$ gives a returned set that
contains $W$ and records $u$.

For each orientation and anchor triple, build the unit-capacity matrix and
the weighted matrices for the Boolean masks, together with their inverses,
once, in $\softO(n^{\omega'})$ time.
The graph remains fixed while targets are processed. By
Proposition~\ref{prop:interface} and its anchor-graph variant, each
target uses $\polylog(n+2)$ full inverse rows or columns and
$\softO(n)$ additional work. Reading those rows or columns from a
stored inverse costs $\softO(n)$, so all targets cost $\softO(n^2)$
after preprocessing. The twenty instances give the stated total bound.
\end{proof}

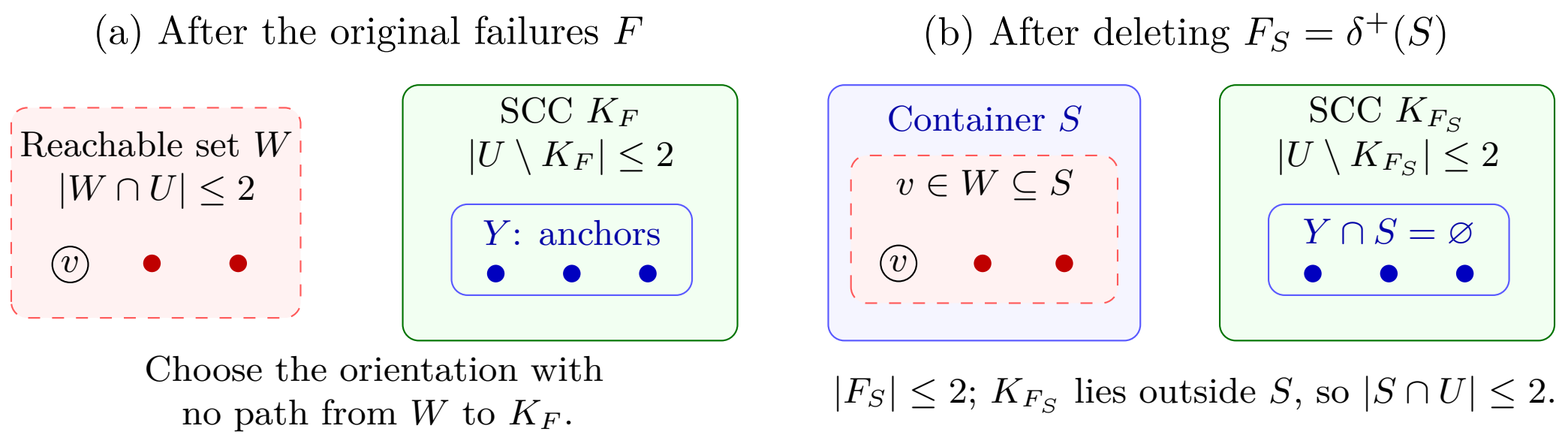
\begin{figure}[t]
\centering
\begin{tikzpicture}[x=1cm,y=1cm,>=Latex,font=\footnotesize]
 \begin{scope}
  \node[font=\small,align=center] at (3.15,3.25)
    {(a) After the original failures $F$};
  \draw[dashed,rounded corners,fill=red!5,draw=red!65] (0,0.75) rectangle (2.55,2.6);
  \node[align=center] at (1.28,2.05)
    {Reachable set $W$\\$|W\cap U|\leq2$};
  \node[draw,circle,inner sep=1pt] at (0.52,1.22) {$v$};
  \fill[red!75!black] (1.24,1.23) circle (2.2pt);
  \fill[red!75!black] (2.0,1.23) circle (2.2pt);
  \draw[rounded corners,fill=green!5,draw=green!45!black] (3.45,0.55) rectangle (6.4,2.8);
  \node[align=center] at (4.93,2.35) {SCC $K_F$\\$|U\setminus K_F|\leq2$};
  \draw[rounded corners,draw=blue!65] (3.88,0.95) rectangle (6.0,1.75);
  \node[blue!65!black] at (4.94,1.5) {$Y$: anchors};
  \foreach \x in {4.27,4.94,5.61} {\fill[blue!75!black] (\x,1.14) circle (2.2pt);}
  \node[align=center,text width=6.4cm] at (3.2,0.08)
    {Choose the orientation with no path from $W$ to $K_F$.};
 \end{scope}
 \begin{scope}[xshift=7.2cm]
  \node[font=\small,align=center] at (3.15,3.25)
    {(b) After deleting $F_S=\delta^+(S)$};
  \draw[rounded corners,fill=blue!4,draw=blue!65] (0,0.55) rectangle (2.75,2.8);
  \node[blue!65!black] at (1.38,2.51) {Container $S$};
  \draw[dashed,rounded corners,fill=red!5,draw=red!65] (0.2,0.88) rectangle (2.55,2.18);
  \node at (1.38,1.92) {$v\in W\subseteq S$};
  \node[draw,circle,inner sep=1pt] at (0.62,1.23) {$v$};
  \fill[red!75!black] (1.36,1.23) circle (2.2pt);
  \fill[red!75!black] (2.08,1.23) circle (2.2pt);
  \draw[rounded corners,fill=green!5,draw=green!45!black] (3.45,0.55) rectangle (6.4,2.8);
  \node[align=center] at (4.93,2.35) {SCC $K_{F_S}$\\$|U\setminus K_{F_S}|\leq2$};
  \draw[rounded corners,draw=blue!65] (3.88,0.95) rectangle (6.0,1.75);
  \node[blue!65!black] at (4.94,1.5) {$Y\cap S=\varnothing$};
  \foreach \x in {4.27,4.94,5.61} {\fill[blue!75!black] (\x,1.14) circle (2.2pt);}
  \node[align=center,text width=6.4cm] at (3.2,0.08)
    {$|F_S|\leq2$; $K_{F_S}$ lies outside $S$, so $|S\cap U|\leq2$.};
 \end{scope}
\end{tikzpicture}
\caption{Five selected terminals give a three-element anchor set outside
$W$. Filled dots show the case in which two selected terminals lie in
$W$; the other three then lie in the indicated large terminal component
in each panel. The components are large in terminal count, not necessarily
in vertex count. The panels use different failure sets; $K_F$ and $K_{F_S}$
need not coincide. Neither $W$ nor $S$ must be an SCC. The algorithm
tries $2\binom53=20$ orientation--anchor pairs, giving $|L_v|\leq40$.}
\label{fig:anchors}
\end{figure}

\begin{corollary}\label{cor:source-bound}
For every strongly connected graph $G$ and nonempty terminal set $U$
satisfying~\eqref{eq:giant}, the sets in~\eqref{eq:true-list} satisfy
$|T_v|\leq40$ for every vertex $v$.
\end{corollary}

\begin{proof}
Run the construction of Lemma~\ref{lem:container} with exact flows.
The coverage argument of Proposition~\ref{prop:lists} uses only its
combinatorial guarantees and gives $T_v\subseteq L_v$, $|L_v|\leq40$.
\end{proof}

\Needspace{22\baselineskip}
\subsection{The recursive algorithm, correctness, and output size}\label{sec:construction}

We now assemble the node construction and its recursive calls.

\begin{algorithm}[Recursive SCC preserver]\label{alg:recursive}
Every pass starts from a separate copy of the original node graph. Temporary
probe graphs used to choose terminals contribute no output edges.

For a strongly connected graph $G$ with $n\geq2$ vertices:
\begin{enumerate}
\item Construct a nonempty terminal set $U$ satisfying terminal
concentration and balance, using Proposition~\ref{prop:terminals}.
Discard the terminal monitor and its probe graphs.
\item Construct the source lists of Proposition~\ref{prop:lists}.
\item Construct the three-root graph $J$ and perform the forward and
reverse list passes of Section~\ref{sec:node-passes}. Let $H_0$ be
the union of their outputs, with reverse edges mapped back.
\item Compute $\SCC(G-U)$ and check balance. Recurse on each induced
graph $G[D]$, and return the union of $H_0$ and the child outputs.
\end{enumerate}
A failed component check reports $\FAIL$. The precise checks and
their error probabilities are given in Section~\ref{sec:probability}.
The one-vertex base case returns no edges; a general initial graph is
handled by the SCC decomposition in Section~\ref{sec:requirements}.
\end{algorithm}

\begin{lemma}\label{lem:preservation}
If the terminal sets, source lists, and local calls are correct at
every node, the recursive output is a two-edge-failure SCC preserver.
\end{lemma}
\begin{proof}
Induct on the number of vertices in a node graph. The one-vertex case
is immediate. Fix a failure set $F\subseteq E(G)$ of size at most two
and an SCC $C$ of $G-F$. If $C$ meets $U$, Lemma~\ref{lem:node-preservation}
preserves it in $H_0-F$.

Otherwise, the paths witnessing strong connectivity of $C$ stay inside
$C$ and avoid $U$. Thus $C$ lies in a single SCC $D$ of $G-U$.
Put $F_D=F\cap E(G[D])$. The set $C$ is an SCC of $G[D]-F_D$:
its internal paths survive, and this subgraph of $G-F$ cannot merge
$C$ with another component. Balance gives $|D|\leq n/2$, so the
inductive hypothesis for the child preserves $C$. Its retained edges
outside $F_D$ also survive $F$.

Every failure SCC is therefore preserved. Since the output is a
subgraph of $G$, distinct failure SCCs cannot be merged. The same
argument applies independently to each original SCC of the input.
\end{proof}

\begin{corollary}\label{cor:global-size}
The recursive construction retains $O(N\log(N+2))$ original edges.
\end{corollary}
\begin{proof}
Node graphs at one depth have disjoint vertex sets, so their sizes sum
to at most $N$. Lemma~\ref{lem:node-size} bounds their total contribution
by $516N$. Balance gives $O(\log(N+2))$ depths.
\end{proof}

\section{Constructing a balanced terminal set}\label{sec:terminals}

This section constructs the terminal set required by
Section~\ref{sec:recursive}. Throughout, $G$ is strongly connected and
has $n\geq2$ vertices. The graph stays fixed while membership in $U$
changes. We seek a nonempty set satisfying terminal concentration
\eqref{eq:giant} and balance~\eqref{eq:balance}.

Condition~\eqref{eq:giant} implies $(2,2)$-unbreakability in HSW's
sense~\cite[Definition 2.1]{hsw}: delete the at most two edges of the
chosen directed cut. No SCC crosses that cut, so the side not containing
the component supplied by~\eqref{eq:giant} has at most two terminals.
HSW's general implication in the other direction allows four terminals
outside one component~\cite[Claim 3.6]{hsw}. We construct the stronger
condition~\eqref{eq:giant} directly.

The exchange pattern is inherited from HSW's balanced hierarchy
construction~\cite[Appendix C.2]{hsw}: remove terminals on
one side of a sparse separation and add boundary endpoints to preserve
balance. Here we use two failed-edge heads and monitor all relevant
separations simultaneously. We first prove the terminal-exchange invariant, then construct the
component records and terminal monitor that implement the exchanges.

\subsection{Terminal exchanges and balance}\label{sec:exchanges}

The construction has two possible outcomes. If some failure set $F$
leaves no vertex-large SCC, let
$L_F=\{\operatorname{head}(e):e\in F\}$ and take $U=L_F$.
This set has at most two vertices, so terminal concentration holds.
Deleting $L_F$ removes every edge of $F$, and hence each SCC of $G-L_F$
lies in an SCC of $G-F$ of size at most $n/2$. Thus balance also holds.
The set $F$ is nonempty because $G$ is strongly connected, so $U$ is
nonempty.

Otherwise, every failure set leaves a vertex-large SCC. Start with
$U=V(G)$. Whenever such a component $C$ of $G-F$ has at least three
terminals outside it, perform the \emph{terminal exchange}
\begin{equation}\label{eq:largeupdate}
 U'=(U\cap C)\cup L_F,
 \qquad L_F=\{\operatorname{head}(e):e\in F\}.
\end{equation}
The exchange keeps terminals in $C$ and adds the failed-edge heads.
Those heads prevent a recursive child from crossing between SCCs of
$G-F$, which is the reason balance survives.

\begin{lemma}\label{lem:terminal-repair}
Starting from a nonempty balanced terminal set, a terminal exchange
on a vertex-large failure SCC preserves balance and nonemptiness.
Over the sequence starting from $U=V(G)$, there are at most $3n$
terminal removals and $2n$ terminal insertions,
counting even an endpoint removed and immediately reinserted.
\end{lemma}
\begin{proof}
The set $C$ is a vertex-large SCC of $G-F$.
Deleting $L_F$ removes $F$, so each SCC of $G-U'$ lies in an SCC of
$G-F$. Outside $C$, these components have fewer than $n/2$ vertices.
Inside $C$, all previous terminals remain deleted; each new SCC is
therefore contained in an SCC of $G-U$. Balance persists.

Let $r_0=|U\setminus C|\geq3$. The terminal exchange removes $r_0$ terminals and
inserts at most two, so its net decrease is at least
$r_0-2\geq r_0/3$. Summing the net decreases, which total at most $n$,
gives at most $3n$ removals. There are at most $n$ terminal exchanges, giving at
most $2n$ insertions. A failure set requiring a terminal exchange is nonempty: for $F=\varnothing$ the only SCC is all of $G$. Thus the inserted endpoint set keeps $U$ nonempty.
\end{proof}

It remains to find a suitable failure set or component efficiently.
The next two subsections represent all vertex-large failure SCCs by
fixed component records. A terminal monitor then finds a record with
at least three terminals outside its component.

\subsection{Representative failure sets and probe graphs}\label{sec:probes}

A one-failure SCC certificate reduces the failure sets to be represented
to a family of size $O(n^2)$. We use the certificate of GIP directly
and prove the representative-family reduction here.

Write $m=|E(G)|$. Compute a one-failure SCC certificate $K\subseteq G$ in $O(m+n)$
time~\cite[Theorem 9.1]{gip}. Its construction uses at most $8n$ edges:
in each orientation, it retains three spanning trees and at most $n-1$
additional edges~\cite[Section 9]{gip}.
Put $\mathcal E=E(K)\cup\{\bot\}$, where $\bot$ means no deletion,
and $F_{f,g}=\{f,g\}\setminus\{\bot\}$ for $f,g\in\mathcal E$.
Repeated entries represent a single failure. Edge identifiers always
refer to the original graph, including when an edge is absent from a
subgraph. Let $Q=|\mathcal E|^2$.

\begin{lemma}\label{lem:failure-family}
Every SCC partition of $G-F$ with $|F|\leq2$ equals the partition of
$G-F_{f,g}$ for some $f,g\in\mathcal E$.
\end{lemma}
\begin{proof}
If $F\subseteq E(K)$, it is represented directly. Otherwise put
$A=F\cap E(K)$, so $|A|\leq1$. Then
$K-A\subseteq G-F\subseteq G-A$. The outer graphs have the same SCC
partition by the one-failure guarantee, so the middle graph has that
partition as well. The set $A$ is represented in the family.
\end{proof}

For these representatives, we need access to every vertex-large
component. A root inside a component makes it recoverable in a probe graph. Sampling a logarithmic number of roots suffices because
each such component occupies more than half the vertices.
For the global confidence parameter $c>0$, sample
\begin{equation}\label{eq:sampling}
 t=\left\lceil\log_2(4Q N^{c+2})\right\rceil
\end{equation}
independent uniform vertices of $V(G)$, with replacement. Call these
vertices \emph{probes}; they are sampled from the fixed vertex set,
not from $U$. For each probe $r$, use Corollary~\ref{cor:rooted} and
Lemma~\ref{lem:batch} to construct a rooted two-failure preserver in
$G$ and another in $G^{\rev}$. Reverse the latter's edges back and let
$Q_r$ be their union. Each $Q_r$ is a subgraph of $G$ with at most $12n$
edges. These constructions precede, and do not use, the terminal set.

With correct local answers, for every original failure set $|F|\leq2$,
\begin{equation}\label{eq:probe-component}
 \text{the SCC of $r$ in $Q_r-F$ equals the SCC of $r$ in $G-F$}.
\end{equation}
Indeed, the forward preserver retains reachability from $r$, and the reverse
preserver retains reachability to $r$. The intersection of these two
reachable vertex sets is the SCC of $r$.
For $x,y$ in this root component, the retained paths
$x\leadsto r\leadsto y$ witness their mutual connectivity. A subgraph
cannot merge distinct original SCCs, so the root component is exact.

\begin{lemma}\label{lem:samples}
Conditional on the node graph, the probes meet every vertex-large SCC
of every $G-F_{f,g}$ except with probability less than
$1/(4N^{c+2})$.
\end{lemma}
\begin{proof}
There is at most one vertex-large SCC per failure set. Each occupies
more than half the vertices, so all $t$ probes miss it with probability
less than $2^{-t}$. A union bound over the at most $Q$ failure sets,
using~\eqref{eq:sampling}, proves the claim.
\end{proof}

We call the event that the probes meet all these vertex-large SCCs the
\emph{probe event}.

\subsection{Component records}\label{sec:records}

Once a first failure is fixed, the second is a one-failure problem in
a sparse graph. We use GIP's dominator characterization directly and
store each resulting component by two subtree roots. This avoids a
separate length-$n$ membership vector for every failure pair.

Fix a probe $r$ and a first failure $f\in\mathcal E$. Compute the SCC
$B_{r,f}$ of $r$ in $Q_r-f$. If $|B_{r,f}|\leq n/2$, discard this pair:
a further deletion cannot make the root component larger. Otherwise
work in the strongly connected graph $(Q_r-f)[B_{r,f}]$.

In a directed graph rooted at $r$, a vertex $x$ \emph{dominates} $v$
if every path from $r$ to $v$ contains $x$. The \emph{dominator tree}
has precisely the dominators of $v$ as the ancestors of $v$, including
$v$ itself. A \emph{flow bridge} is an edge $(x,y)$ that belongs to
every path from $r$ to its head $y$. Its deletion makes $y$ unreachable.
Compute the forward and reverse dominator
trees, and their flow bridges, in $\softO(n)$ time on this sparse
graph~\cite[Section 2.1]{gip}.

For $g=(x,y)\in E(K)$, let $A_g\subseteq B_{r,f}$ be the vertex set of the forward dominator subtree
rooted at $y$ if $g$ is a forward flow bridge, and let $R_g\subseteq B_{r,f}$ be the vertex set of the
reverse dominator subtree rooted at $x$ if $g$ is a reverse flow bridge.
An absent edge, $g=\bot$, or a nonbridge gives an empty set in the
corresponding tree. In particular, $g=f$ gives two empty sets.

\begin{lemma}[Root component; cf.~{\cite[Theorem 3.7]{gip}}]
\label{lem:root-component}
The SCC of $r$ in $Q_r-F_{f,g}$ is
\begin{equation}\label{eq:root-component}
 C_{r,f,g}=B_{r,f}\setminus(A_g\cup R_g).
\end{equation}
\end{lemma}
\begin{proof}
Paths between vertices of $B_{r,f}$ stay inside that SCC. In its induced
graph, deleting a forward flow bridge makes exactly its head's dominator
subtree unreachable from $r$; the reverse tree gives the vertices that
lose reachability to $r$~\cite[Theorem 3.7]{gip}. A vertex belongs to
the remaining root SCC precisely when neither loss occurs.
\end{proof}

Store a record for $(r,f,g)$ only when $|C_{r,f,g}|>n/2$. Mark its
failure set $F_{f,g}$ as \emph{covered}. Each record stores $g$ and the
two subtree roots, not the vertex set $C_{r,f,g}$. Subtree sizes and
intersection sizes determine its size. Appendix~\ref{sec:monitor}
computes all these intersection sizes in $O(n\log n)$ time for a fixed
$(r,f)$. There are $O(nt)$ such pairs, each with $O(n)$ candidate second
failures. The total preprocessing time and space are $\softO(n^2)$.

If some failure set $F$ is uncovered, compute $\SCC(G-F)$ exactly.
If it has a vertex-large SCC, report $\FAIL$. Otherwise return
$U=L_F$, where $L_F=\{\operatorname{head}(e):e\in F\}$.
Deleting $L_F$ removes all failed edges, so balance holds. Also
$|L_F|\leq2$, which implies the terminal concentration property. This set is nonempty,
since $F=\varnothing$ would leave the strongly connected $G$ vertex-large.
This branch always terminates; it makes only one SCC scan of the
original graph. On the probe event of Lemma~\ref{lem:samples} and
correct rooted constructions, it cannot report $\FAIL$.

\subsection{The terminal monitor}\label{sec:terminal-monitor}

The recorded vertex sets are fixed. The remaining task is to detect
when one has too many terminals outside it, without rescanning its
vertices after each change to $U$. Inclusion--exclusion reduces the
count to membership in the two dominator subtrees.

Assume every failure set is covered. Initialize $U=V(G)$. For each
stored record maintain the integer
\begin{equation}\label{eq:monitor-count}
 \mu_{r,f,g}=|U\setminus C_{r,f,g}|
 =|U\setminus B_{r,f}|+|U\cap A_g|+|U\cap R_g|-|U\cap A_g\cap R_g|.
\end{equation}
The subtraction counts a terminal in both subtrees only once. A
\emph{membership change} inserts one vertex into $U$ or removes one
vertex from $U$. The graph and both trees remain fixed.

\begin{lemma}\label{lem:monitor}
After $\softO(n^2)$ preprocessing, the records occupy $\softO(n^2)$
space and support a membership change in $\softO(n^{3/2})$ time.
They find a record with $\mu_{r,f,g}\geq3$, or establish that no such
record exists, in $\softO(n)$ time.
\end{lemma}
\begin{proof}
For a fixed $(r,f)$, represent each record by the preorder positions of
its two subtree roots. A position of zero denotes an empty subtree.
A vertex whose membership changes belongs to a subtree exactly when the subtree root
is its ancestor. A heavy-path decomposition~\cite{st} writes each
root-to-vertex path as $O(\log n)$ preorder intervals. Thus
\eqref{eq:monitor-count} changes by $O(\log^2 n)$ signed rectangle
additions on the $O(n)$ record points, or by a shared offset if the
vertex is outside $B_{r,f}$.

Appendix~\ref{sec:monitor} gives an $O(n)$-space structure with
$O(\sqrt n\log n)$ time per rectangle addition and for reporting a record
of maximum weight.
Across $O(nt)$ pairs, each membership change therefore costs
$\softO(n^{3/2})$. Store each pair's maximum and a record attaining it.
Scanning these $O(nt)$ maxima finds a violation in $\softO(n)$ time.
All rectangles of one membership change are applied before a maximum
is queried. The preprocessing and storage bounds include the two trees,
their decompositions, and all record points for every pair.
\end{proof}

\subsection{The terminal-set construction and its cost}\label{sec:terminal-algorithm}

Construct the representative failure family, the probe graphs, and the
component records. The uncovered-failure branch in Section~\ref{sec:records}
either returns a terminal set of size at most two or reports $\FAIL$.
If every representative is covered, initialize $U=V(G)$ and the terminal
monitor, then perform terminal exchanges as follows.

A record is \emph{violating} if $\mu_{r,f,g}\geq3$. Such a record identifies
both the component to keep terminals in and the failed-edge endpoints that
will preserve balance. For a violating record, put $F=F_{f,g}$ and recover
$C=C_{r,f,g}$ by an SCC search in the sparse $Q_r-F$, in $O(n)$ time.
Perform the terminal exchange~\eqref{eq:largeupdate}.
Apply the corresponding membership changes to the terminal monitor. Repeat
until all record counts are at most two. Finally check that $U$ is
nonempty and compute $\SCC(G-U)$ to verify balance; report $\FAIL$ if
either check fails. Each terminal exchange is also required to decrease $|U|$.

\begin{proposition}\label{prop:terminals}
If the probe event of Lemma~\ref{lem:samples} holds, all local answers are
correct, and the maintained matrices are nonsingular, the procedure constructs a
nonempty $U$ satisfying~\eqref{eq:giant} and~\eqref{eq:balance} in
$\softO(n^a)$ time and $\softO(n^2)$ space.
\end{proposition}
\begin{proof}
The uncovered-failure branch has already been justified. Otherwise
balance holds initially and persists by Lemma~\ref{lem:terminal-repair}.
At termination every represented failure set has a component with at
most two terminals outside it. Lemma~\ref{lem:failure-family} extends
this guarantee to every original failure set, proving the terminal
concentration property. The final checks therefore succeed.

The $2t$ rooted passes take $\softO(n^a)$ time by Lemma~\ref{lem:batch}.
The certificate, record construction, and SCC computations in the original node graph take
$\softO(n^2)$ time. Lemmas~\ref{lem:monitor} and
\ref{lem:terminal-repair} bound all monitor updates by
$\softO(n^{5/2})$. The $O(n)$ searches for violating records and SCC
searches in the probe graphs take $\softO(n^2)$ time. Since $a>5/2$, the total is
$\softO(n^a)$. The probe graphs have $O(nt)$ edges in total, and the
monitors and one active algebraic pass occupy $\softO(n^2)$ space.
Here and throughout the paper, soft-$O$ includes powers of $\log(N+2)$.
\end{proof}

\paragraph{Validity of a returned terminal set.} The algorithm checks that each $Q_r$ is a subgraph of $G$ with at
most $12n$ edges. This check, together with exact SCC computations,
certifies any terminal set that the stage returns. Each monitored SCC then lies inside
an SCC of $G-F$. Coverage and counts at most two imply the terminal
concentration property, while the final exact check establishes balance. Thus any
terminal set returned by this stage is valid. Correct rooted constructions
and the probe event ensure that the stage returns a set rather than $\FAIL$.
The later algebraic edge-selection stages retain the Monte Carlo
guarantee stated in Theorem~\ref{thm:main}.

\section{The matrix data structure and the running time}\label{sec:algebra}

The graph construction uses local calls to select incoming edges, then
applies incoming restrictions as a pass proceeds. This section implements
those operations. We first encode bounded-flow questions by determinants,
then show how inverse queries recover cut sides and original boundary
edges. Finally, epochs support the changing graph efficiently, and we
combine the costs and error probabilities to prove Theorem~\ref{thm:main}.

\subsection{Matrix encoding and computational conventions}\label{sec:model}

The matrix represents degree-constrained edge subsets of a query graph.
Its determinant will test whether such a subset exists; a flow-threshold
gadget translates the flow questions into these degree constraints.
We use the common-basis encoding of two partition-matroid unions from
CKS~\cite[Section 3.1]{cks}. Partitioning the
edges by tail imposes an outgoing-degree bound, and partitioning them
by head imposes an incoming-degree bound. A common basis of the two
$K$-fold unions selects exactly $K$ edges in every part. We state the
specialization explicitly to fix the matrix blocks and the changes
needed by our queries. CKS use randomized linear representations of
matroid unions due to Narayanan, Saran, and Vazirani~\cite{nsv}.
Testing for a common basis by a determinant is classical;
Harvey~\cite[Section 4.3]{harvey} describes this approach.

Let $B$ be the graph on which a flow is queried. Write $\nu=|V(B)|$;
$\nu=n$ for a pass and $\nu=n+1$ when an anchor sink is present.
A capacity-$q$ edge represents $q$ distinct unit copies. Copies share
the original edge identifier when they encode an original edge, so
deleting that original edge removes all its copies. Copies are not
separate possible failures.

Fix $K=3$ for unit flow queries or $K=9$ for weighted queries. Add two
\emph{encoding vertices} $a,b$ and $K$ distinct \emph{padding loops} at
every vertex. A padding loop is an auxiliary unit self-loop: selecting
it contributes one incoming and one outgoing edge at the same vertex.
It will fill unused degree. The vertices $a,b$ and these loops belong
only to the matrix encoding; they never belong to the cut sides or
membership vectors of $B$.

We first encode a degree constraint. A \emph{$K$-regular edge subset}
selects exactly $K$ incoming and $K$ outgoing unit edges at each encoded
vertex, counting a selected loop in both degrees. For each unit edge
$e=(u,v)$, introduce a scalar variable $x_e$ and two $K$-vectors
$\alpha_e,\beta_e$ of variables. All these variables are distinct. Define
a matrix with $(\nu+2)\times(\nu+2)$ vertex blocks, each of size
$K\times K$, by
\begin{equation}\label{eq:matrix}
 M[u,v]=\sum_{e=(u,v)}x_e\alpha_e\beta_e^{\transpose}.
\end{equation}
Its scalar dimension is $D=K(\nu+2)$: a scalar row or column is
indexed by a vertex together with a coordinate in $\{1,\ldots,K\}$.
For example, $K=3$ gives a $3\times3$ block per ordered vertex pair.
The extra coordinates encode degree constraints. The sum includes
padding and other auxiliary edges when present; each edge contributes
a rank-one block.

\begin{lemma}[CKS partition-matroid specialization]\label{lem:det}
The polynomial $\det M$ is nonzero if and only if the encoded graph has
a $K$-regular edge subset. Its total degree is at most $3D$.
\end{lemma}

\begin{proof}[Encoding identity]
Write $M=P\operatorname{diag}(x)Q^{\transpose}$, where the column for
$e=(u,v)$ places $\alpha_e$ in the tail block of $P$ and $\beta_e$ in
the head block of $Q$. Cauchy--Binet expands $\det M$ over subsets of
$D$ edges. A term can be nonzero exactly when the subset has $K$
incoming and $K$ outgoing edges at every vertex. Distinct edge subsets
have distinct monomials in the $x_e$, so those terms cannot cancel.
Appendix~\ref{sec:primitive-details} gives the block calculation,
which also establishes validity in characteristic two.
\end{proof}

The implementation accumulates each unit edge's constant-size block
directly into $M$, in $O(n^2)$ time and space. It does not materialize
the potentially much wider matrices $P,Q$, which serve only in the proof.

The \emph{maintained matrix} encodes the current graph with all $K$ padding
loops at every vertex; $a,b$ have only their padding loops. The loops alone
form a regular subset, so the determinant is a nonzero polynomial after
any permanent original-edge restrictions. A \emph{query matrix} temporarily
modifies this encoding to test a flow threshold.

An \emph{encoding instance} contains the matrices for one pass or for one
orientation and anchor triple in the source-list construction. Assign each
variable an independent uniform element of $\F$ and store each evaluated
maintained matrix, its determinant, and its inverse. The assignments stay
fixed throughout the instance, including all epochs of a pass. Different
instances use independent assignments. A singular maintained matrix causes $\FAIL$.

\paragraph{Matrix algorithms and exponent notation.}
The exponent $\omega'$ fixed after Theorem~\ref{thm:main} is realized
by a fixed recursive exact bilinear algorithm over $\basefield$.
A bilinear scheme multiplies pairs of linear forms, one in each input
matrix, and combines these products linearly to obtain the output.
Its rank is the number of these products. Following the bilinear-rank
definition~\cite[Definition 5.1]{blaser}, $\omega$ is the infimum of
$\beta$ for which $t\times t$ matrix multiplication has exact schemes
of rank $O(t^\beta)$. For every fixed $\omega'>\omega$, choose a
sufficiently large constant-size scheme and apply it recursively to
matrix blocks. This gives $O(t^{\omega'})$ operations. Its coefficients and recursion rule
are independent of the graph size, and it works over every extension
of $\basefield$. Evaluating recursive subproducts sequentially uses
$O(t^2)$ space for $t\times t$ matrices.
A suitable constant-size scheme can be found by exhaustive search:
enumerate block sizes and coefficient arrays over $\basefield$, and
verify the bilinear identities and the resulting exponent bound.
For a fixed rational exponent threshold above $\omega$, this search
terminates and has cost independent of $N$.

Exact determinant computation and inversion reduce to matrix
multiplication, with $\softO(t^{\omega'})$ field operations and
$\softO(t^2)$ space~\cite{bh,jps}. Pivoting selects nonzero field
entries; no ordering or approximate arithmetic is needed. The matrices
we invert have nonzero determinant polynomials, though a random
evaluation can still be singular. Section~\ref{sec:probability}
accounts for this event. Cauchy--Binet, the determinant lemma, and the
Woodbury identity used below are exact identities over the field.

\paragraph{Finite-field representation and bit cost.}
The algorithm evaluates its matrices over a polynomial-size extension
$\F$ of the fixed base field $\basefield$. In the standard
polynomial-basis representation~\cite[Chapters 17, 19, and 20]{shoup}, an element
has $O_c(\log(N+2))$ base-field coefficients, so arithmetic is exact
and has polylogarithmic bit cost. Appendix~\ref{sec:field} chooses the
extension degree and gives its construction, uniform sampling, and
error-probability bounds. One concrete instantiation uses $\basefield=\mathbb F_2$
and Strassen's algorithm~\cite{strassen}.

Combinatorial $O$-bounds count operations and storage words on a RAM
with $O(\log N)$-bit words. All soft-$O$ bounds include field arithmetic
and storage, suppressing powers of the global $\log(N+2)$ even at
small recursion nodes. Thus the same node routine can be used at every
size $n\geq2$.

\subsection{Flow tests and recovery of original boundary edges}\label{sec:flow-tests}

A local call needs a capped flow value, a farthest cut side when the
value is one or two, and the original edges on that side's boundary.
The determinant encoding supplies these answers through the following
temporary modifications to the matrix encoding. The current graph is unchanged.

To test $\flow_B(s,t)\geq\mu$, where $1\leq\mu\leq K$, temporarily
remove $\mu$ padding loops at each of $a,b$ and insert $\mu$ distinct
copies of each of
\[
 a\longrightarrow s,\qquad t\longrightarrow b,\qquad b\longrightarrow a.
\]
The return edges $b\to a$ close the desired $s$--$t$ flow into a
circulation. The degree constraints then force the required amount
through $s$ and $t$. This is a circulation variant of CKS's bounded-flow
reduction~\cite[Section 4.1]{cks}. We verify the variant because its
padding loops ensure that the determinant of every maintained matrix is a
nonzero polynomial.

\begin{lemma}\label{lem:threshold}
The query determinant is a nonzero polynomial exactly when
$\flow_B(s,t)\geq\mu$ before the encoding vertices and gadget are added.
\end{lemma}

\begin{proof}
At $a$ and $b$, each available incoming and outgoing count is exactly
$K$. A regular subset must therefore select all gadget edges and all
remaining loops there. After removing the gadget and all padding loops
from that subset, the selected edges in $B$ have outgoing minus incoming
degree $\mu$ at $s$, $-\mu$ at $t$, and zero elsewhere. They give a
capacity-feasible integral flow of value $\mu$, possibly with cycles.

Conversely, take $\mu$ edge-disjoint simple $s$--$t$ paths among the
unit copies of $B$. Add the gadget. Incoming and outgoing degrees now
agree and are at most $\mu\leq K$ at every vertex. Select enough padding
loops at each vertex to bring both degrees to $K$.
Lemma~\ref{lem:det} proves the equivalence.
\end{proof}

The threshold test returns a Boolean value: true exactly when the
evaluated determinant is nonzero. A nonzero answer certifies the
threshold, but a zero answer can be erroneous. The integer capped flow
value is the number of true tests among thresholds $1,\ldots,K$; it is
not the field value of a determinant. Query matrices are never inverted.
Section~\ref{sec:probability} bounds errors in all threshold tests and
evaluations of maintained matrices, including adaptive queries.

\paragraph{Recovering a farthest cut side.}

Suppose $B=X+A$, with at most two fresh unit shortcuts in $A$, all from
$s$, and $\flow_B(s,t)=\lambda\in\{1,2\}$. Lemma~\ref{lem:farthest}
gives each bit of the farthest-side membership vector $C$: temporarily
add a unit edge $s\to w$ and test whether the value is still $\lambda$.
Set the bits for $s,t$ to one and zero, respectively.

An \emph{outgoing star} $S(s,C;q)$ means the auxiliary edges
$s\to w$ for all $w\in C\setminus\{s\}$, each of capacity $q$. The
edges are fresh, even if $X$ or $A$ already has the same endpoints.
We use this star to force a weighted minimum cut to contain the already
known set $C$, and then read a small count from its value.

Fix a Boolean mask $h:E(X)\to\{0,1\}$ on the fixed original edge
identifiers. The \emph{masked cut-value query} gives each original edge
$e$ capacity $3+h(e)$, gives each shortcut capacity three, and adds
$S(s,C;9)$. Its inputs are $X,s,t,A,\lambda,C,h$ as above; its output
is the following integer.

\begin{lemma}\label{lem:force}
The unique minimum source side of this weighted graph is $C$, and its
cut value is
\begin{equation}\label{eq:maskvalue}
 3\lambda+\sum_{e\in\delta_X^+(C)}h(e)\leq8.
\end{equation}
\end{lemma}

\begin{proof}
The star is internal to $C$, which has the displayed value. A source
side omitting a vertex of $C$ cuts a star edge and has value at least
nine. A proper superset of $C$ has at least $\lambda+1$ unit boundary
edges in $B$, by maximality of $C$. Its weighted value is at least
$3(\lambda+1)>3\lambda+2$. These cover all competing source sides.
Thus the multiplier three exceeds the largest possible mask perturbation
of two, and the star capacity nine exceeds the candidate value of eight.
\end{proof}

At each recursion node, assign its $m$ original edges distinct local
IDs $0,\ldots,m-1$, with a table to input edge IDs. These IDs remain
fixed after restrictions. Individual bit sums need not determine two
IDs: the pairs $\{00,11\}$ and $\{01,10\}$ have the same sum in each
position. Products of two bits reveal which bits belong to the same
edge. Put $L=\lceil\log_2\max\{2,m\}\rceil$ and prepare the
$O(\log^2(n+2))$ masks
\[
 h_i(e)=\operatorname{bit}_i(e),\qquad
 h_{ij}(e)=\operatorname{bit}_i(e)\operatorname{bit}_j(e)
 \quad(0\leq i,j<L).
\]
Bits refer to the ordinary binary expansion of the integer local ID.
Each mask has its own maintained $K=9$ matrix, with original capacities
$3+h(e)$. These are the \emph{mask encodings}; the $K=3$ matrix for
unit-capacity queries is the \emph{unit encoding}. We build these matrices
once per encoding instance and maintain them under permanent restrictions. A
masked query uses the relevant existing matrix and adds only its
temporary shortcuts, star, and threshold gadget. It does not rescan
or reweight all original edges. Thresholds $1,\ldots,9$ return its
integer cut value; subtract $3\lambda$ to get the mask sum.

Inspect $A$ directly to find its crossing edges. The number of unknown
original boundary edges is $r=\lambda-|\delta_A^+(C)|\in\{0,1,2\}$.
If $r=0$, nothing remains. If $r=1$, the single-bit sums give the ID.
If $r=2$, write these sums as $a_i\in\{0,1,2\}$. Distinct IDs have a
bit $j$ with $a_j=1$. The sum $a_{ij}$ for mask $h_{ij}$ is bit $i$ of
the edge whose bit $j$ is one. The other edge has bit $a_i-a_{ij}$.
Thus $O(\log(n+2))$ masked values recover both IDs.

If a count lies outside its stated range, no separating bit exists, or
a decoded ID is invalid, report $\FAIL$. Check that the decoded edges
are present, distinct, cross $C$, and match all queried sums. These
checks take $\softO(n)$ time using the membership vector and ID table.

\subsection{Inverse queries and incoming-edge selection}\label{sec:inverse-queries}

The determinant lemma is a standard identity, also used by CKS's
sensitivity framework~\cite[Appendix A]{cks}. The cost here depends on
the support of each temporary change. Low rank alone does not give a
linear-time query when both factors are dense.

Let $Z=M^{-1}$ be the inverse of the current maintained matrix, of dimension $D$.
For a set $R$ of vertex blocks, $E_R$ is the $D\times K|R|$ matrix of
identity columns at their coordinates. An outgoing star, the shortcuts,
and the threshold gadget change only block rows in $R=\{s,t,a,b\}$.
Write their change as $E_RD_R$, where $D_R$ is the $K|R|\times D$
array of row differences. The determinant identity is
\begin{equation}\label{eq:rowlemma}
 \det(M+E_RD_R)=\det(M)\det(I+D_RZE_R).
\end{equation}
The final determinant has constant dimension. Given the constant number
of full columns $ZE_R$, constructing the star and testing its thresholds
takes $O(n)$ further operations. For cut memberships there is no star:
each added edge changes only a constant number of entries. All $\nu$
membership tests reuse the same full columns and take $O(n)$ further
work in total. They are not $\nu$ separate requests for full inverse data.

Deleting a group of original edges entering $t$ changes its column
block. With the shortcuts and threshold gadget, the temporary change is
\begin{equation}\label{eq:mixed}
 T E_t^{\transpose}+E_RD_R.
\end{equation}
Here $T$ has dimensions $D\times K$ and can be dense, whereas $D_R$
has only constantly many nonzero entries. There is no star in this
query. Apply the determinant identity with left factor $[T\ E_R]$ and
right factor $[E_t\ D_R^{\transpose}]^{\transpose}$. The small matrix
uses the four products
\[
 E_t^{\transpose}ZT,\quad E_t^{\transpose}ZE_R,
 \quad D_RZT,\quad D_RZE_R.
\]
Obtain the rows of $Z$ selected by $E_t^{\transpose}$ and by the nonzero
columns of $D_R$. There are constantly many. Their products with $T$
cost $O(n)$, and the other products cost no more. Thus a group-deletion
threshold test needs full inverse rows; a masked cut-value query needs full
inverse columns. No permitted query has both a dense row change and a
dense column change. This restriction avoids $\Theta(n^2)$ work per test.

\paragraph{Selecting three incoming edges.}

A positive test for flow at least three certifies existence but does not identify
edges to retain. We recover three final edges by eliminating groups
of incoming edges while keeping that threshold true.

At a leaf of $\Local$ with flow at least three, order the current incoming original
edges of $t$ by ID and recursively split the list into halves. This gives
a binary tree of depth $O(\log(n+2))$. Maintain a set of tentatively
deleted edges, initially empty. At a visited tree node, test whether
deleting its group together with the accepted tentative deletions leaves
flow at least three in $X+A$. If so, accept that deletion. Otherwise
leave the group present and process its children; a rejected singleton
is retained. Only original edges entering $t$ are tested, and no shortcut
enters $t$. Each test returns a Boolean threshold answer.

With correct answers, a rejected singleton remains necessary after all
later tentative deletions: further deletions cannot raise the flow
without that edge. The final retained set $K_t$ is therefore
inclusion-minimal for retaining a flow of value at least three. Three edge-disjoint paths use only
three final edges, so retaining those edges already suffices. Minimality
then gives $|K_t|=3$. Every rejected group meets $K_t$: otherwise the final
supporting graph would be contained in the rejected trial graph. Hence
at most three groups at each depth are rejected and only $O(\log(n+2))$
groups are tested. Each test has form~\eqref{eq:mixed} and costs $O(n)$
after the required full rows of the inverse are available. The current graph is
unchanged until the caller applies an incoming restriction.

\begin{proposition}\label{prop:interface}
Assume all threshold answers are correct and the maintained matrices
are nonsingular.
Given the current unit and mask encodings, one top-level call to $\Local$
uses $\polylog(n+2)$ requests, each for
constantly many full rows or columns of an inverse, and $\softO(n)$
further work. It returns at most six original incoming
edge IDs with the guarantee of Lemma~\ref{lem:local}. The encodings
consist of one unit matrix and $O(\log^2(n+2))$ mask matrices.
An observed shortcut must raise the capped flow by one, until it reaches
three; otherwise the procedure reports $\FAIL$. This enforces the
two-shortcut recursion limit.
\end{proposition}

\paragraph{The anchor-graph variant.}
The flow tests, cut-membership tests, and recovery of original edges from
cut boundaries of size at most two also apply
when $X$ has a sink $z$ and three fixed unit copies of $y\to z$ for
each anchor $y$. The sink belongs to the queried graph, unlike the
encoding vertices $a,b$. In each weighted encoding give each anchor copy
capacity three and mask zero, for total capacity nine per anchor. A cut
to $z$ of unit value at most two excludes every anchor, so its boundary
contains no anchor edge. The weighted candidate value is at most eight;
other cut sides containing an anchor cost at least nine. The preceding
membership and recovery proofs therefore apply. Replace $\nu$ by $n+1$
in the matrix dimensions. We use only these operations at target $z$;
we do not apply the guarantee for selecting original incoming edges there.

\subsection{Matrix passes and epochs}\label{sec:batch}

The query interface requests rows and columns of the current inverse;
this section implements those requests and applies permanent incoming
restrictions. A restriction can delete $\Theta(n)$ edges at one target,
so its cost must depend on the sparse replacement block. The batching
argument uses the standard low-rank-update and periodic-rebuilding
framework for dynamic matrix inverses~\cite[Section 2]{bns}. The
special property used here is that each target's replacement block is
sparse and its columns have not been changed earlier in the pass.

A pass processes each original target once. It can issue
$\polylog(n+2)$ inverse row or column requests at each target. It then
restricts that target's incoming original edges. Let $d_v$ be the retained
indegree, and suppose
\begin{equation}\label{eq:degreebudget}
 \sum_v d_v=O(n).
\end{equation}
Protected edges, if any, count toward $d_v$. The bound applies to the sum;
individual $d_v$ may be larger than a constant. The data structure supports
initialization, full inverse-row or inverse-column retrieval, one target
restriction, and rebuilding after $b$ restrictions. Lemma~\ref{lem:node-size} verifies these conditions for the rooted
passes and list passes in the recursive construction.

\Needspace{10\baselineskip}
\begin{lemma}\label{lem:batch}
Assume the query answers are correct and all maintained matrices
are nonsingular. A pass satisfying the stated bounds on inverse queries
and total retained indegree
with the interface of Proposition~\ref{prop:interface} takes
\begin{equation}\label{eq:pass}
 \softO\left(\frac{n^{\omega'+1}}{b}+n^2b+n^2\right)
\end{equation}
time and $\softO(n^2)$ space, for any integer $1\leq b\leq n$.
\end{lemma}

\begin{proof}
Consider one encoding of dimension $D=O(n)$. At the start of an epoch,
compute its matrix $M_0$, inverse $Z=M_0^{-1}$, and determinant. An epoch
contains at most $b$ target restrictions. With $r=O(b)$ changed scalar
columns, write
\[
 M=M_0+\Delta E^{\transpose},\qquad P=Z\Delta,
 \qquad \Sigma=I_r+E^{\transpose}P.
\]
Here $E$ is the $D\times r$ matrix of identity columns at changed
coordinates, $\Delta$ is their $D\times r$ array of differences from
$M_0$, $P$ is $D\times r$, and $\Sigma$ is $r\times r$.

Before target $v$ is processed, its columns still equal those of $M_0$.
Let $Q_v$ be their $D\times K$ replacement. They have nonzero entries
in at most $1+d_v$ vertex blocks: the padding block and the retained
incoming tails. With $E_v$ selecting these target coordinates,
\begin{equation}\label{eq:sparseupdate}
 Z(Q_v-M_0E_v)=ZQ_v-E_v.
\end{equation}
Multiplying $Z$ by the replacement block $Q_v$ forms the appended columns of $P$
in $O(n(1+d_v))$ operations. Their total cost is $O(n^2)$ by
\eqref{eq:degreebudget}. Forming and storing the corresponding differences
$\Delta$ also fits this bound.

For this restriction, append $E_v$ to $E$, append
$Q_v-M_0E_v$ to $\Delta$, and append $ZQ_v-E_v$ to $P$.
The matrix $\Sigma$ consequently gains $K$ rows and columns.
The standard Schur-complement identity updates its inverse and determinant.
The complement is nonsingular whenever the old
and new maintained matrices are nonsingular. The added block has constant
size, so forming its complement and the inverse-update products costs
$O(b^2)$. The determinant of the complement multiplies the old determinant. For the
initially empty $\Sigma$, take determinant one and initialize from the
first constant-size block.

The inverse update identity (Woodbury) gives
\begin{equation}\label{eq:woodbury}
 M^{-1}=Z-P\Sigma^{-1}E^{\transpose}Z,
 \qquad \det M=\det M_0\det\Sigma.
\end{equation}
To get column $j$, first read $y=E^{\transpose}Z_{[:,j]}$, then compute
$w=\Sigma^{-1}y$, then $Z_{[:,j]}-Pw$. These cost $O(b)$, $O(b^2)$,
and $O(nb)$, respectively. To get row $i$, first compute
$w=P_{[i,:]}\Sigma^{-1}$ and then $Z_{[i,:]}-wE^{\transpose}Z$; the
costs are $O(b^2)$ and $O(nb)$. Thus constantly many requested full rows
or columns cost $O(nb+b^2)$. The pass has $\softO(n)$ such requests.
Further query work is $\softO(n^2)$ by Proposition~\ref{prop:interface}.
Maintaining $\Sigma^{-1}$ costs $O(nb^2)\leq O(n^2b)$ in total.

There are $O(n/b)$ epochs, each with $\softO(n^{\omega'})$ rebuilding cost.
This proves~\eqref{eq:pass}. The unit and mask encodings add logarithmic
factors. Their $D\times D$ matrices and $D\times O(b)$ update arrays
occupy $\softO(n^2)$ space. A singular maintained matrix or update
complement causes $\FAIL$.
\end{proof}

Set
\begin{equation}\label{eq:parameters}
 b=\left\lceil n^{(\omega'-1)/2}\right\rceil,
 \qquad a=\frac{\omega'+3}{2}.
\end{equation}
Then $5/2<a<3$, and the pass costs
$\softO(n^a)$. Corollary~\ref{cor:rooted} uses one such pass and satisfies
the bound on total retained indegree. Lemma~\ref{lem:node-size} gives the same bound for the list
passes. All rooted passes used for probes satisfy it as well.

\subsection{Global running time and space}\label{sec:global}

Assume correct query answers and the probe event at every node.
For a node with $n\geq2$ vertices, the costs are
\begin{center}
\begin{tabular}{ll}
\toprule
Stage & Time\\
\midrule
Balanced terminal set & $\softO(n^a)$\\
Source lists & $\softO(n^{\omega'})$\\
Three-root graph $J$ & $\softO(n^a)$\\
Forward and reverse list passes & $\softO(n^a)$\\
SCC decomposition and other work & $\softO(n^2)$\\
\bottomrule
\end{tabular}
\end{center}
Constructing the terminal set includes $O(\log N)$ rooted passes and
$\softO(n^{5/2})$ monitor work. Since $\omega'<a$ and $5/2<a$, the
static list computations and monitor do not dominate. Thus the node
cost is $O(n^a\polylog(N+2))$, with a logarithmic exponent independent
of the node. For its disjoint children,
\begin{equation}\label{eq:recursion}
 \sum_i n_i^a
 \leq (n/2)^{a-1}\sum_i n_i
 \leq \frac{n^a}{2^{a-1}}.
\end{equation}
Summing this geometric decrease gives $O(n^a\polylog(N+2))$
time for a recursion tree rooted at an $n$-vertex SCC. If the original
SCCs have sizes $N_j$, then $\sum_jN_j^a\leq N^a$, so the entire
recursion forest takes $O(N^a\polylog(N+2))$ time. The global logarithmic factor covers field
arithmetic, probe counts, and mask encodings at every node size.
Because $\omega'<\omega+2\varepsilon$, this implies the time bound
in Theorem~\ref{thm:main}.

Corollary~\ref{cor:global-size} gives the output-size bound.

Process children sequentially. Discard the parent's inverse matrices
and terminal-monitor tables before entering a child. Parent graphs and pending child
data can be stored within $\softO(N^2)$ space; their sizes decrease
geometrically along a recursion path. The output itself uses
$O(N\log(N+2))$ space. This proves the space bound.

\subsection{Success probability}\label{sec:probability}

There are three substantive random choices. Uniform vertex probes
expose vertex-large failure components. Random coefficients evaluate
the determinant encodings, using the CKS framework and the classical
Schwartz--Zippel bound~\cite{cks,schwartz,zippel}. A standard randomized
finite-field construction uses an exact irreducibility
test~\cite{rabin,shoup}. Bounded rejection sampling implements the
uniform draws. Selected terminal roots and anchor triples are
deterministic; the failure sets to be preserved are universally
quantified.

Field construction and sampling can report $\FAIL$. The terminal stage
checks the conditions on any $U$ it returns. In later edge selection,
however, a nonzero determinant polynomial may evaluate to zero without
triggering a check, so a returned subgraph can be incorrect. This is the
Monte Carlo error covered below. All arithmetic on sampled coefficients
is exact. The proof must account for reuse of those coefficients and
for queries chosen from earlier answers.

First assume exact combinatorial answers to every flow query. At each
randomized node, sample fresh probes from its fixed vertex set.
Lemma~\ref{lem:samples} holds conditional on all prior choices. Each
node either removes a nonempty terminal set before recursing or is a
one-vertex leaf. Charging the node to a removed or leaf vertex gives
at most $N$ nodes; the looser bound $2N$ suffices here. The conditional
union bound gives total probe error less than
$2N/(4N^{c+2})\leq\tfrac14N^{-c}$ for $N\geq2$.
The confidence parameter in~\eqref{eq:sampling} uses the original $N$,
not the local $n$.

Next fix the probe random tapes and consider the \emph{ideal execution}
with exact flow answers. Fix all tie rules. Graph choices, target groups,
and masks in this execution then depend on the graph and these tapes,
not on private field values. Each original or auxiliary edge role receives independent coefficients
when first used, and repeated uses of that role reuse those coefficients.
The assignment is fixed throughout a pass, including its epochs.
Different encoding instances use independent assignments.
Appendix~\ref{sec:field} specifies the roles and shows that their tables
occupy $\softO(n^2)$ space. This lets all ideal query determinants be
viewed as polynomials in one fixed set of independent variables.

Each determinant that should be nonzero on this ideal execution is a
nonzero polynomial of degree $O(N)$ by Lemmas~\ref{lem:det} and
\ref{lem:threshold}. This includes maintained matrices. There are
polynomially many such tests and rebuilds when the probe event holds at every node.
The Schwartz--Zippel bound~\cite{schwartz,zippel} states that a nonzero
polynomial of total degree $d_0$, evaluated at independent uniform field
elements, vanishes with probability at most $d_0/|\F|$. Applying it
to these polynomials and taking a union bound makes their total error
less than $\tfrac14N^{-c}$ over the field in~\eqref{eq:concrete-field}. Appendix~\ref{sec:field}
gives an explicit polynomial bound on the number of tests and hence a
sufficient extension degree.
Determinants that are identically zero cannot give false positive
threshold answers.

This argument covers adaptivity. Until the first field error, the actual
execution follows the ideal one. If the first error occurs, one of the
ideal nonzero polynomials evaluated to zero. No independence between
successive query answers is assumed. The choices of pivots during matrix inversion
are not used to choose graph operations.

The algorithm reports $\FAIL$ on a singular maintained matrix or update
complement, nonmonotone threshold answers, an impossible capped value,
or a failed decoding or progress check. In particular, a leaf selection
must return exactly three edges, a local call at most six, and a source
list at most 40; otherwise it reports $\FAIL$. For a
decoded boundary list, it checks that the reported identifiers are
distinct, present in the current graph, cross the reported side, and
match all queried bit sums. It also checks the required terminal progress.
These checks use the stored membership vectors and edge table; they do
not scan all edges to verify that no boundary edge was omitted. Their
cost fits the charged $\softO(n)$ query work or the charged SCC scan.
These checks detect some algebraic errors; a false zero at a true flow
threshold can remain undetected. The probability bound covers both
detected and undetected errors.
Sampling and evaluation errors together have probability at most
$\tfrac12N^{-c}$.
On their complement, all graph and matrix arguments above apply.

The field and its construction are specified in
Section~\ref{sec:model} and Appendix~\ref{sec:field}. Field setup fails
with probability at most $\tfrac14N^{-c}$. Bounded rejection sampling of
uniform integers contributes at most another $\tfrac14N^{-c}$, as detailed
in that appendix. The total failure probability is less than $N^{-c}$.
Every execution terminates: terminal exchanges must decrease $|U|$,
recursion proceeds only after the balance check, and all sampling and
evaluation loops have prescribed limits on their numbers of trials and tests.
When the probe event holds at every node, field construction and sampling
succeed, and no field-evaluation error occurs, none of these checks reports $\FAIL$,
and the algorithm returns the preserver
within the stated bounds. Each field operation has polylogarithmic
bit cost. This proves Theorem~\ref{thm:main}.

\section{Conclusion}

The lower bound and the HSW upper bound establish optimal
size $\Theta_k(N\log N)$ in the worst case for every fixed $k\geq2$. For two failures,
the construction achieves this size in randomized subcubic time by
combining terminal concentration, source lists, and matrix passes.

The construction leaves open faster algorithms at the optimal size and
an extension of the subcubic bound to larger fixed $k$.
Our source-list construction uses the fact that one shortcut raises a
minimum cut of value one to two. A comparable implementation for larger
values of $k$ would need to control both the number of required sources and
the cost of recovering their cut boundaries. Further questions concern
the optimal size when $k$ grows with $N$, and the size
and construction time on restricted graph classes, such as graphs
without strong bridges.

\section*{Acknowledgment of AI Assistance}

ChatGPT (OpenAI) was used to assist with the development and
verification of parts of the proofs, as well as to improve the
writing and presentation of the paper.

\newcommand{\etalchar}[1]{$^{#1}$}

\appendix
\section{A deterministic baseline from prior results}\label{sec:baseline}

Protect a linear-size one-failure certificate $K$, then delete every
other edge whose removal preserves all required SCC partitions.
Protecting $K$ reduces verification to failure sets drawn from its
$O(n)$ edges. At termination, no further edge outside $K$ is removable.
The HSW existence theorem then forces this minimal remainder to have
$O(n\log(n+2))$ edges. The proof below implements the verification and
establishes this size bound.

\begin{proposition}[Elementary baseline from prior results]
\label{prop:baseline}
The one-failure certificate algorithm of~\cite{gip} and the size
theorem of~\cite{hsw} imply a deterministic $O(n^5)$-time construction
of an $O(n\log(n+2))$-edge two-failure SCC preserver on a simple graph.
\end{proposition}

\begin{proof}
Construct a one-failure SCC preserver $K\subseteq G$ with $O(n)$ edges.
Keep all of $K$ throughout the algorithm. For any intermediate graph
$K\subseteq X\subseteq G$ and any $|F|\leq2$ with $F\nsubseteq E(K)$,
put $A=F\cap E(K)$. Then $|A|\leq1$ and
\[
 K-A\ \subseteq\ X-F\ \subseteq\ G-F\ \subseteq\ G-A.
\]
The first and last graphs have the same SCC partition. Every intermediate
graph has that partition as well. Thus only the $O(n^2)$ sets
$F\subseteq E(K)$ of size at most two need to be checked.

Represent an SCC partition by an array that assigns each vertex the
smallest vertex identifier in its component. Two such arrays are equal
exactly when the partitions are equal, and comparison takes $O(n)$ time.
For each $f\in E(K)\cup\{\bot\}$, construct a one-failure certificate
$C_f$ of $X-f$, where $\bot$ means no deletion. Each certificate has
$O(n)$ edges and takes $O(m+n)$ time. For each
$g\in E(K)\cup\{\bot\}$, an ordinary SCC search on $C_f-g$
returns the partition of $X-(\{f,g\}\setminus\{\bot\})$;
a missing edge has no effect. The certificates cost
$O(n(m+n))$ to construct. The $O(n^2)$ SCC searches and partition
comparisons cost $O(n^3)$. Hence a complete check of $X$ takes $O(n^3)$.
Store the corresponding reference table for $G$ once.

Start with $X=G$. Process each edge $e$ outside $K$ once. Apply the
preceding test to the candidate $X-e$, comparing every partition to
the stored reference for $G$. Set $X\leftarrow X-e$ exactly when all
comparisons succeed. This keeps $X$ a two-failure preserver of
$G$. If an edge is rejected, some failure set witnesses a split when that
edge is omitted. Further safe deletions preserve the reference SCC, and
cannot restore the missing path without the edge. The rejection therefore
remains valid. There are at most $m=O(n^2)$ checks, giving $O(n^5)$ time.

Let $H$ be the final graph. The HSW size theorem~\cite[Theorem 4.2]{hsw}, applied to $H$, supplies
some two-failure preserver $D\subseteq H$ with $O(n\log(n+2))$ edges.
We use its existence only; the algorithm need not construct $D$.
The graph $K\cup D$ also preserves $H$, and hence preserves $G$.
If an edge of $H$ lay outside $K\cup D$, it could still be deleted safely.
This contradicts the final rejection property. Therefore
\[
 |E(H)|\leq |E(K)|+|E(D)|=O(n\log(n+2)).
\]
\end{proof}

\section{Finite-field construction and probability accounting}\label{sec:field}

There are polynomially many determinant tests, each of degree $O(N)$.
Schwartz--Zippel therefore makes a polynomial-size field sufficient:
the total error is at most the sum of these degrees divided by the
field size. We specify coefficient reuse first, bound that sum, and
then construct the extension and its exact uniform sampler.
The base field $\basefield$ is fixed by the matrix multiplication
algorithm, and $q_0$ is a constant. If $q_0=p_0^e$ is not prime, fix
once and for all an irreducible degree-$e$ polynomial over
$\mathbb F_{p_0}$ and use its quotient representation for $\basefield$.
Its operation tables and bit costs are constants independent of $N$.

\subsection{Coefficient roles and reuse}

An \emph{edge role} specifies an encoding instance, an encoding, an edge type,
endpoints, and a unit-copy index. For each role, assign independent variables
to $x$ and to each coordinate of $\alpha$ and $\beta$. Edge types distinguish
original edges, fixed anchor edges, shortcuts, edges for cut-membership tests, stars,
threshold gadgets, and padding loops. An original type also records its
fixed edge ID. Repeated queries of one role reuse its variables; parallel
copies and distinct types use different variables. For shortcuts, also
record an occurrence number in $\{1,2\}$; each occurrence has one unit
copy in the unit encoding and three in a weighted encoding. Edges used for cut-membership tests use a separate type. All other multiplicities are at most nine.
There are $O(n^2)$ possible edge roles
per encoding in an instance. In particular, roles have no query-number
index that would create a new stored set for every test.

This full assignment is only a device for the probability proof.
The implementation assigns a role's variables when it first needs them.
Different encoding instances use separate roles. It discards their
tables when the instance ends. Within a pass, the coefficient tables stay
fixed across epochs. At any time, the current encodings and their role
tables use $\softO(n^2)$ space. Creating the coefficients for a query
costs at most its already charged $\softO(n)$ work.

\subsection{A sufficient field size}

Fix probe choices for which the probe event holds at every node of the
ideal execution from Section~\ref{sec:probability}.
Put $x=N+2$, $\tau=\log_2x$, and define
\[
 B_c=10^{12}(c+20).
\]
There are at most $N$ nodes; we use $2N$ as a loose upper bound. Since $|E(K)|\leq8n$,
\eqref{eq:sampling} gives $t\leq(c+20)\tau$ probes per randomized node.
Each such node has at most twenty orientation--anchor instances and
$2t+8$ edge-selection passes: two per probe, at most six for the final
terminal roots, and two for the source lists.

Count every threshold test, including tests of identically zero
polynomials, and every nonsingularity check for a maintained matrix at initialization,
permanent update, or epoch rebuild. These are observable operations,
so their total can be monitored without knowing the correct answers.
The following bounds include all unit and mask encodings:
\begin{center}
\begin{tabular}{ll}
\toprule
Operation & Tests and nonsingularity checks\\
\midrule
One orientation--anchor instance, for all sources & $10^5x^3$\\
One edge-selection pass, for all targets & $10^8x^3$\\
\bottomrule
\end{tabular}
\end{center}
To check these loose constants, the local edge-ID length is at most
$3\tau$, so the number of encodings is at most $13\tau^2$. A source in an
anchor instance has at most two stages with flow value one or two.
A top-level local call has at most four recursive invocations, at most two
with flow value one or two, and at most two leaves with flow at least three.
Each stage with flow value one or two uses at most $N+1$
membership tests and at most $13\tau^2$ mask queries, each with at most
nine thresholds. Each leaf with flow at least three tests at most
$1+6\lceil\log_2x\rceil$ groups. Thus $10^4x^2$ threshold tests
suffice per source or top-level local call. A pass has at most $40N$
such calls. Its matrix nonsingularity checks add at most $100x^3$: each of
the $13\tau^2$ encodings has at most $N$ updates and $N+1$ initializations
or rebuilds. These estimates imply both table entries.

Summing over source-list instances and all passes gives at most
\[
 2N\bigl(20\cdot10^5+(2(c+20)\tau+8)10^8\bigr)x^3
 \leq B_cx^6,
\]
using $N\leq x$, $\tau\leq x$, and $N\geq2$.
The algorithm counts these operations and reports $\FAIL$ before
exceeding $B_cx^6$.

The small determinants used to evaluate queries can be rational
functions of the assigned variables, because they contain the inverse of
a maintained matrix. Their zero tests are equivalent to zero tests of the full
query determinants whenever the maintained matrix is nonsingular.
Likewise, let $H$ be the constant-size Schur complement used to append
a target block in Lemma~\ref{lem:batch}. It satisfies
\[
 \det H=\frac{\det M_{\rm new}}{\det M_{\rm old}}.
\]
Thus its nonsingularity follows from the nonsingularity of the two successive
maintained matrices, whose checks are already counted. The probability bound applies to full
matrix determinant polynomials, rather than to rational expressions
in intermediate inverses.

Each relevant polynomial has degree at most $27(N+3)$: the largest
encoding has $K=9$, and an anchor graph adds at most one ordinary vertex
before the two encoding vertices. Choose
$A_c=81B_c$. The sum of all relevant degrees is then at most
$A_c(N+2)^7$. Choose the extension
\begin{equation}\label{eq:concrete-field}
 \F=\basefield[T]/(f(T)),\qquad
 r=\deg f=\left\lceil\log_{q_0}
       \bigl(4A_c(N+2)^{c+12}\bigr)\right\rceil,
\end{equation}
where $f$ is monic and irreducible over $\basefield$.
This field has size at least $4A_c(N+2)^{c+12}$ and less than $q_0$
times that quantity.
Schwartz--Zippel~\cite{schwartz,zippel} bounds the total field-evaluation error by
\[
 \frac{A_c(N+2)^7}
 {4A_c(N+2)^{c+12}}
 \leq \frac14 N^{-c}.
\]
This extension-size choice, together with the polynomial-basis arithmetic
below, gives the bit-cost implementation used in the main theorem.

\subsection{Finding and representing the extension}

Sample the $r$ lower coefficients of a monic degree-$r$ polynomial
$f\in\basefield[T]$ independently and uniformly. Test irreducibility
by the following exact criterion:
\begin{align*}
 T^{q_0^r}&\equiv T\pmod f,\\
 \gcd\bigl(f,T^{q_0^{r/h}}-T\bigr)&=1
       &&\text{for every prime divisor $h$ of $r$}.
\end{align*}
Rabin's criterion is exact~\cite{rabin}. For fixed $q_0$, trial division
of $r$, modular powering, and polynomial gcds give an $O(r^4)$
base-field-operation test. At least a $1/(2r)$ fraction of monic
degree-$r$ polynomials are irreducible for $r\geq2$; for $r=1$ they
all are. We use these standard facts and the resulting randomized
construction from~\cite[Section 20.1]{shoup}.

Try at most
\[
 L=\left\lceil2r\bigl((c+2)\ln(N+2)+\ln4\bigr)\right\rceil
\]
independent candidates. Report $\FAIL$ if none is irreducible. The chance
of this event is at most
$\exp(-L/(2r))\leq\tfrac14(N+2)^{-c-2}<\tfrac14N^{-c}$.
The total setup cost is polynomial in $\log(N+2)$.
After success, every field operation is exact. Represent an element by
its $r$ coefficients modulo $f$. Schoolbook multiplication and reduction,
and the polynomial extended Euclidean algorithm for a nonzero inverse,
use $O(r^2)$ base-field operations. Coefficientwise sampling gives exactly
uniform field elements. In the binary instance, these are uniform $r$-bit
strings.

\subsection{Implementing uniform sampling with random bits}

For a uniform integer in $\{0,\ldots,h-1\}$, use
$\lceil\log_2 h\rceil$ fair bits, reject values at least $h$, and repeat;
for $h=1$ return zero. Each trial succeeds with probability greater than
$1/2$. A probe is obtained by one uniform integer in
$\{0,\ldots,n-1\}$, interpreted as an index in the node's vertex array.
Fresh probes use fresh bits; repeated vertices are allowed. The
three-terminal anchor sets are enumerated deterministically and require
no random draws.

Use the same integer sampler for coefficients in the fixed base field.
We give a separate bound for the number of integer draws, including the
$r$ coefficients of each extension-field element. With $x,\tau$ as above,
put
\[
 \kappa=1+\log_2(4A_c)+c+12,
 \qquad r\leq\kappa\tau.
\]
Each initialized encoding has fewer than $10^4x^2$ scalar variables:
there are at most nine edge-type and shortcut-occurrence combinations,
at most nine unit copies per endpoint pair and combination, and at most
$2K+1\leq19$ variables per copy. There are at most $N+3\leq5x/4$
encoded vertices. The initialization count is at most
$B_cx^6$, and roles are reused within each instance. Hence
matrix coefficients use at most $10^4B_c\kappa x^9$ base-field
integer draws. Vertex probes use at most $2N(c+20)\tau\leq2(c+20)x^2$ draws.
Field setup uses at most $Lr\leq3\kappa^2(c+3)\tau^3$ draws.
Consequently, the constant
\[
 D_c=10^4B_c\kappa+2(c+20)
              +3\kappa^2(c+3)+1
\]
bounds the total by $D_cx^{10}$ on every ideal execution in which the
probe event holds at every node.
Impose this total draw limit, and allow each draw at most
\[
 J=\left\lceil\log_2(4D_c)+(c+12)\log_2(N+2)\right\rceil
\]
trials. Report $\FAIL$ before exceeding the total draw limit, or if all
permitted trials for a draw are rejected. A union bound over
the draws gives rejection-abort probability at most
$D_cx^{10}2^{-J}\leq\tfrac14x^{-c-2}<\tfrac14N^{-c}$.
Conditional on acceptance, each draw is exactly uniform. Couple the
bounded procedure with exact rejection sampling using the same bits;
their executions agree until the first abort. This justifies applying
the uniform-sample analysis above and adding the abort probability.
Each draw has polylogarithmic bit cost.

\Needspace{14\baselineskip}
\section{Implementing the terminal monitor}\label{sec:monitor}

We give the data structures used in Lemma~\ref{lem:monitor}. All trees
and record coordinates are fixed while $U$ changes. Work first with one
pair $(r,f)$ and write $B=B_{r,f}$, $b=|B|$. The forward and reverse
dominator trees have vertex set $B$. A subtree includes its root.
Fix a heavy-path decomposition~\cite{st} of each tree and number its
vertices in a depth-first preorder that visits the heavy child first.
Both the static intersection counts and the dynamic updates use
these fixed preorder numbers.
Two different point sets serve different purposes: vertex points count
intersections of fixed subtrees, while record points identify the
component counts affected by a change to $U$.
\begin{center}
\begin{tabularx}{\textwidth}{@{}p{0.19\textwidth}p{0.36\textwidth}X@{}}
\toprule
Point type & Two coordinates & Operation\\
\midrule
Vertex $v$ & Preorder positions of $v$ in the two trees & Static rectangle
counts for subtree intersections\\[2pt]
Record $g$ & Preorder positions of the roots of $A_g$ and $R_g$ & Dynamic
rectangle additions to record weights\\
\bottomrule
\end{tabularx}
\end{center}
The invariant is that each record's weight plus its pair's common
offset equals $\mu_{r,f,g}$, the number of terminals outside its component.

\subsection{Subtree intersections and initial counts}

In this preorder, a vertex receives its number
when first visited, before its children. A subtree is an interval
$[\operatorname{in}(v),\operatorname{out}(v)]$, where
$\operatorname{out}(v)$ is the largest number in the subtree.
Represent each vertex $v\in B$ by its two preorder numbers. The size
of an intersection of a forward and a reverse subtree is the number
of these points in the corresponding rectangle.

All $O(n)$ required counts can be computed in $O(n\log n)$ time and
$O(n)$ space by an offline sweep. For a rectangle
$[x_1,x_2]\times[y_1,y_2]$, create one query at $x_2$ and a subtractive
query at $x_1-1$. Sort the vertices and queries by first coordinate.
Insert vertices whose first coordinate is at most the query coordinate,
then count inserted second coordinates in $[y_1,y_2]$. A Fenwick tree~\cite{fenwick} supports insertion and interval counts
in $O(\log n)$ time and linear space. Empty subtrees have intersection
count zero.
Thus we obtain $|A_g\cap R_g|$, and hence $|C_{r,f,g}|$, for all $g$.

Keep only records with $|C_{r,f,g}|>n/2$. Initially $U=V(G)$, so the
record's weight is $|A_g|+|R_g|-|A_g\cap R_g|$, and the common offset
is $n-|B|$. Their sum is $\mu_{r,f,g}$. Only subtree roots, sizes,
weights, and coordinates are retained; no component membership vector
is stored for each second failure.

\subsection{A membership change gives signed rectangle additions}

Each subtree is an interval, and each root-to-vertex path
is a union of $O(\log n)$ disjoint intervals that can be enumerated in
that time. These standard properties let a membership change address
all relevant subtree roots at once.

For a stored record, let $(x_g,y_g)$ be the preorder positions of the
roots of $A_g$ and $R_g$, using zero for an empty subtree. A membership
change at vertex $v$ has sign $\Delta=1$ for insertion and $-1$ for
removal. If $v\notin B$, add $\Delta$ to the common offset. Otherwise
let $I_1,\ldots,I_p$ and $J_1,\ldots,J_q$ be the ancestor-path intervals
of $v$ in the two trees. Perform the rectangle additions
\[
 \begin{array}{ll}
 +\Delta&\text{on each }I_i\times[0,b],\\
 +\Delta&\text{on each }[0,b]\times J_j,\\
 -\Delta&\text{on each }I_i\times J_j.
 \end{array}
\]
Ancestor intervals use only coordinates $1,\ldots,b$, so they exclude
the coordinate zero reserved for an empty subtree. A point receives $\Delta$ precisely when
$v\in A_g\cup R_g$, proving~\eqref{eq:monitor-count}. Both signs and
overlapping forward and reverse subtrees are handled by this identity.
Apply the complete collection before testing a maximum.

\subsection{Rectangle additions and maximum-weight queries}

For $s>0$ record points, sort by first coordinate and partition the
order into blocks of at most $h=\lceil\sqrt{s}\rceil$ points. Within
each block sort by second coordinate. Store their weights in a segment tree with lazy range additions,
maintaining the maximum and an index attaining it. A range addition
takes $O(\log(s+2))$ time, and the root reports the block maximum.

For a rectangle $[x_1,x_2]\times[y_1,y_2]$, its first-coordinate range
selects a contiguous portion of the global sorted order, with at most
two partial blocks. In each full block, binary searches find the
second-coordinate interval, and one range addition updates it. In each
partial block, inspect its at most $h$ points and perform single-point
updates where needed. There are $O(\sqrt{s})$ blocks, so the total cost
is $O(\sqrt{s}\log(s+2))$. Equal coordinates are retained as separate
points; inclusive lower and upper searches keep this bound unchanged.
Scan the block maxima after the update to retain the maximum for the
whole pair $(r,f)$, adding its common offset. A pair with no records
is omitted. The space is $O(s)$, and the bound also covers $s=1$.

For every pair $(r,f)$, both dominator trees, the heavy-path data, and
its $O(n)$ records occupy $O(n)$ words. Store membership in $B$ and
both preorder numbers in arrays indexed by the node's vertex identifiers;
these also take $O(n)$ words per pair. There are $O(nt)$ pairs.
Updating all pairs after one membership change takes
$O(nt\sqrt n\,\log^3(n+2))=\softO(n^{3/2})$ time. Scanning their
stored maxima costs $O(nt)=\softO(n)$. This proves the bounds used in
Lemma~\ref{lem:monitor}, including recovery of a witnessing failure
set and probe graph.

\section{Details of the local and algebraic primitives}\label{sec:primitive-details}

The main text states the guarantees and their sources. This appendix
records the short adaptations needed for the auxiliary-edge and
matrix conventions used here.

\subsection{Correctness of the six-edge local call}

The following proves Lemma~\ref{lem:local} using BCR's farthest-cut
selection argument~\cite[Lemma 13]{bcr}.

\begin{proof}
The head $h$ of a cut edge lies outside $C$. Lemma~\ref{lem:farthest}
shows that each new shortcut raises the flow by one. Thus a call with flow
value one has at most one child, with flow value two. A call with flow value
two has at most two children, each with flow value three. There are at most
four recursive invocations and at most two shortcuts on a branch. A leaf
with flow at least three retains three edges. A direct
edge into $t$ replaces a recursive branch and cannot increase this bound.
The total is at most six.

For correctness, consider any call reached during
$\Local(X,s,t)$. Write $B=X+A$ for its augmented graph. Every shortcut
in $A$ leaves $s$, and none enters $t$. Let $\delta_X^-(t)$ denote the
original edges of $X$ entering $t$, and let $S\subseteq\delta_X^-(t)$
be the returned set. The invariant is that, for every
$F\subseteq E(X)$ with $|F|\leq2$,
\[
 \begin{aligned}
 &s\leadsto t\text{ in }(X+A)-F\\
 &\qquad\Longrightarrow\quad
 s\leadsto t\text{ in }
 \bigl(X-(\delta_X^-(t)\setminus S)+A\bigr)-F.
 \end{aligned}
\]
Failures affect only original edges; the shortcuts in $A$ remain available.
We establish the invariant by induction on the remaining recursion depth.
If the flow is at least three, take three edge-disjoint
paths. At most two of them contain failed original edges. The three final
edges are original, since no shortcut enters $t$. Retaining these final
edges leaves a surviving path.

For a low-flow call, take a surviving simple $s$--$t$ path $P$ in $B-F$.
Let $(u,h)$ be its first edge that leaves $C$. If $h=t$, the retained edge
and the prefix of $P$ give the required path. Otherwise the child has
the path formed by its new shortcut and the suffix $P[h,t]$. Induction
preserves reachability in that child after its incoming restriction.
If a resulting path uses the new shortcut, replace it by $P[s,h]$.
This prefix survives $F$ and never enters $t$. It is unaffected by an
incoming restriction at $t$. The replacement gives a walk, and hence a
path, in the parent's restricted graph. The union of child choices can
only add edges. This proves the inductive step. The zero-flow case is vacuous.
At the initial call, $A=\varnothing$, giving the guarantee of
Lemma~\ref{lem:local}.
\end{proof}

\subsection{The partition-matroid determinant}

For completeness, the specialization of CKS's common-basis
polynomial~\cite[Section 3.1]{cks} in Lemma~\ref{lem:det} is as follows.

\begin{proof}
Let $\widetilde m$ be the number of encoded unit edges. Form
$D\times\widetilde m$ matrices $P,Q$. Column $e$ of $P$ contains
$\alpha_e$ in the block for the tail of $e$ and zero elsewhere; column
$e$ of $Q$ contains $\beta_e$ in its head block. Then
$M=P\operatorname{diag}(x)Q^{\transpose}$. The Cauchy--Binet identity gives
\[
 \det M=\sum_{\substack{S\subseteq E_{\rm enc}\\|S|=D}}
       \det(P_{[:,S]})\det(Q_{[:,S]})\prod_{e\in S}x_e.
\]
Here $E_{\rm enc}$ is the set of encoded unit edges, and $P_{[:,S]}$
is the square matrix of columns indexed by $S$, in fixed edge order;
the same order is used for $Q$.

The first selected matrix is singular unless exactly $K$ selected
edges leave every vertex: otherwise some row block has fewer than $K$
columns that can support it. With exactly $K$ at every vertex, a column
permutation makes it block diagonal with $K\times K$ blocks of distinct
variables, each with nonzero determinant. The second selected matrix
has the corresponding incoming condition. For a regular subset their
product is a nonzero polynomial. Different subsets have different
monomials in the separate variables $x_e$, so these terms cannot cancel.
Within each block, different determinant permutations also have distinct
monomials. This remains true in characteristic two. Every entry of $M$
has degree three, giving degree at most $3D$.
\end{proof}

\end{document}